# Finding Natural, Dense, and Stable Frustrated Lewis Pairs on Wurtzite Crystal Surfaces

Xi-Yang Yu,[a] Zheng-Qing Huang,[a] Tao Ban,[a] Yun-Hua Xu,[b] Zhong-Wen Liu,*[c] and Chun-Ran Chang*[a,b]

[a] X.-Y, Yu, Dr. Z.-Q., Huang, T. Ban, Prof. Dr. C.-R., Chang
Shaanxi Key Laboratory of Energy Chemical Process Intensification
School of Chemical Engineering and Technology, Xi'an Jiaotong University
Xi'an 710049 (China)
E-mail: changcr@mail.xjtu.edu.cn

[b] Prof. Dr. Y.-H., Xu, Prof. Dr. C.-R., Chang
Shaanxi Key Laboratory of Low Metamorphic Coal Clean Utilization
School of Chemistry and Chemical Engineering, Yulin University
Yulin 719000 (China)

[c] Prof. Dr. Z.-W., Liu
Key Laboratory of Syngas Conversion of Shaanxi Province
School of Chemistry and Chemical Engineering, Shaanxi Normal University
Xi'an 710119 (China)
E-mail: zwliu@snnu.edu.cn

**Abstract:** The surface frustrated Lewis pairs (SFLPs) open up new opportunities for substituting noble metals in the activation and conversion of stable molecules. However, the applications of SFLPs on a larger scale are impeded by the complex construction process, low surface density, and sensitivity to the reaction environment. Herein, wurtzite-structured crystals such as GaN, ZnO, and AlP are found for developing natural, dense, and stable SFLPs. It is revealed that the SFLPs can naturally exist on the (100) and (110) surfaces of wurtzite-structured crystals. All the surface cations and anions serve as the Lewis acid and Lewis base in SFLPs, respectively, contributing to the surface density of SFLPs as high as $7.26 \times 10^{14}$ cm$^{-2}$. Ab initio molecular dynamics simulations indicate that the SFLPs can keep stable under high temperatures and the reaction atmospheres of CO and $H_2O$. Moreover, outstanding performance for activating the given small molecules is achieved on these natural SFLPs, which originates from the optimal orbital overlap between SFLPs and small molecules. Overall, these findings not only provide a simple method to obtain dense and stable SFLPs but also unfold the nature of SFLPs toward the facile activation of small molecules.

## Introduction

Since Douglas W. Stephan *et al.*[1] creatively proposed sterically encumbered phosphines and boranes for reversible hydrogen activation, named as frustrated Lewis pairs (FLPs), the concept of FLPs has been extended to the activation of a range of small molecules, including $CO_2$, $NH_3$, olefins, and alkynes.[2] For instance, the intramolecular FLPs, $o$-$C_6H_4(NMe_2)(B(C_6F_5)_2)$, can achieve high chemo- and stereoselective hydrogenation of internal alkynes into cis-alkenes under mild conditions.[2b] Though the activation and conversion of small molecules can be greatly improved by using FLPs, the costly recycling of soluble FLPs still limits their large-scale applications.[3] Therefore, developing surface FLPs (SFLPs) based on solid materials is a promising direction to overcome the inherent disadvantages of traditional homogeneous FLPs.[4]

To date, several strategies, such as doping heteroatoms, introducing foreign groups, and creating surface vacancies, have been proposed to construct SFLPs.[4,5] Chen *et al.*[6] demonstrated that SFLPs could be constructed by doping B/Al atoms on two-dimensional black phosphorous, which were very active for hydrogen dissociation. On hydroxylated metal oxide surfaces, the oxygen atoms in hydroxyl groups and their nearby metal atoms can cooperatively form SFLPs.[7] Similarly, SFLPs can also be constructed by introducing Lewis pair-functional groups on metal-organic frameworks (MOF).[8] The MOF-based FLPs possess excellent performance in selectively photocatalytic hydrogenation of $CO_2$, but the reactivity can only be maintained for a few cycles.[8c] Without introducing heteroatoms or foreign groups, our group found that regulating surface oxygen vacancies on metal oxides could also construct SFLPs.[9] By creating appropriate oxygen vacancies on $CeO_2$(110) and (100), the SFLPs consisting of Ce cation as Lewis acid and its nearby oxygen anion as Lewis base can be successfully constructed and were revealed to possess excellent performance in the activation of $H_2$, $CH_4$, and CO. Moreover, the experimental studies verify that the $CeO_2$ nanorod with massive oxygen vacancies can effectively catalyze the hydrogenation of alkenes and alkynes, which is mainly attributed to the high activity of SFLPs on $CeO_2$ for $H_2$ dissociation.[9a] Overall, though significant progress in SFLPs design and applications has been achieved in recent years, the strategies for constructing SFLPs are not simple and efficient enough for large-scale applications.

The difficulties of constructing SFLPs not only lie in creating active sites at a precise level of sub-nanometer but also result from the segregation of acid and base sites to avoid chemical bond formation.[4] For example, the MOF-based SFLPs were prepared by introducing foreign acid and base groups to the designed anchoring sites for artificially separating the Lewis acid and Lewis base sites,[8a] which undergoes three complex steps and cannot guarantee the uniformity of the prepared SFLPs. Another drawback of the constructed SFLPs is that the surface density is too low to meet the requirements of practical applications. As mentioned above, the amount of SFLPs depends on a few doped atoms, surface vacancies, or introduced foreign groups, which results in the low surface density of SFLPs and thus the low overall activity of SFLP catalysts. In addition, the artificial SFLPs are not stable enough under reaction conditions and atmospheres. According to the literature, the state-of-the-art SFLPs can only maintain activity for one hundred hours or even several cycles,[10] which may be caused by the agglomeration of the doped metal atoms for SFLPs or the poisoning of active sites by reaction atmospheres such as CO and $H_2O$. Considering the shortages of the artificially constructed SFLPs above, it is important to pursue an efficient method to develop SFLPs with high surface density and long-term stability to push forward the large-scale applications of SFLPs.

In this work, we report that wurtzite-structured crystals, such as GaN, ZnO, and AlP, are ideal candidates for developing natural SFLPs. It is found that the SFLPs can naturally exist on the (100) and (110) surfaces of wurtzite-structured crystals without any elaborate surface engineering. As all the surface atoms can serve as frustrated Lewis acid/base sites, the surface density of natural SFLPs can reach up to $7.26 \times 10^{14}$ cm$^{-2}$. Importantly, the SFLPs can remain stable under high temperatures and in atmospheres of CO and $H_2O$. The natural SFLPs on wurtzite crystal structures are demonstrated with remarkable performance in the activation of small molecules, which originated from the desirable small molecule···SFLPs orbitals overlap. This work not only offers a simple strategy for constructing natural, dense, and stable SFLPs but also unravels the nature of SFLPs for small molecule activation.

## Results and Discussion

**Finding natural SFLPs over wurtzite crystal surfaces**

To explore the existing possibility of natural SFLPs, fourteen crystal structures, i.e., fluoride, rocksalt, rutile, anti-fluoride, zincblende, corundum, anatase, wurtzite, bixbyite, cuprite, quartz, A-type $M_2O_3$ (M represent metal), B-type $M_2O_3$, and CuO-type structures (see details in Table S1), were screened and analyzed in the following discussion (Figure 1). According to the definition of FLPs,[11] a cation (labeled as $LA_1$) and its next-nearest-neighbor anions (labeled as $LB_2$) are likely to form FLPs ($LA_1$···$LB_2$ in Figure 1a). However, the generation of FLPs may face hindrances caused by the neighboring atoms between $LA_1$ and $LB_2$. In the previous studies of SFLPs construction on ceria surfaces, the nearest oxygen atoms (like $LB_1$ in Figure 1a) of Ce atoms were removed to create SFLPs active sites on $CeO_2$ (100) and (110) surfaces, whereas the SFLPs were still unable to form on the $CeO_2$(111) surface by removing the nearest oxygen atoms due to the hindrance caused by the nearest-neighbor cations (like $LA_2$ in Figure 1a).[9a] Therefore, in the following discussion, the

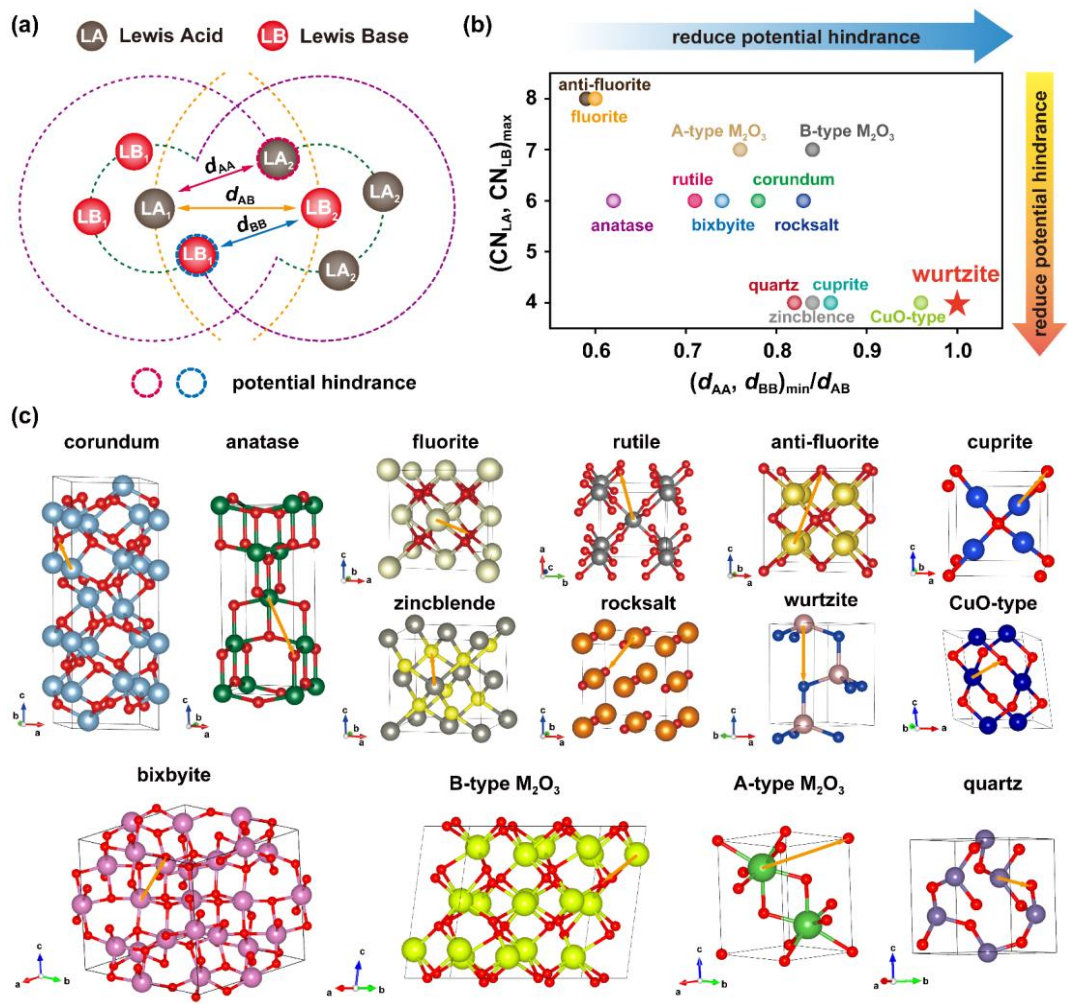

**Figure 1.** Crystal structure analysis. (a) Schematic image of promising FLPs and its potential hindrance. (b) Analysis of bulk structures through established criteria. (c) Schematic image of bulk structures of the fourteen crystals. The spheres with large radii and small radii represent cations and anions, respectively. The orange arrows in (c) illustrate potential FLPs.

Lewis sites and their coordination environment are analyzed to evaluate the possibility of FLPs naturally existing. In general, a cation ($LA_1$) and its nearest-neighbor anions ($LB_1$) usually form chemical bonds and thus are named classical Lewis pairs (CLPs). Though the coordination number (CN) of $LA_1$ cannot be completely eliminated, minimizing the coordination numbers of $LA_1$, namely removing the anions $LB_1$, is a feasible pathway to reduce the hindrance caused by the $LB_1$. Similarly, the CN of $LB_2$ should also be as small as possible to allow the $LA_1 \cdots LB_2$ to satisfy the concept of FLPs. Hence, the first criterion for screening crystal is the maximum value of CN of both $LA_1$ and $LB_2$ (denoted as $(CN_{LA}, CN_{LB})_{max}$). In addition to the CN of $LA_1$, the nearest-neighbor cation ($LA_2$) may lead to hindrance. The $LA_2$ can lead to potential hindrance only if the $LA_2$ approaches to the $LA_1 \cdots LB_2$ pairs; that is when $LA_2$ forms a chemical bond with $LB_2$. On crystal surfaces, as the ratio of $LA_1 \cdots LA_2$ distance (denoted as $d_{AA}$) to $LA_1 \cdots LB_2$ distance (denoted as $d_{AB}$) increases, the potential hindrance caused by $LA_2$ gradually decreases (see details in Figures S1–S2). Likewise, it's expected to get a large $d_{BB}/d_{AB}$ value as large as possible. Therefore, the second screening principle is the ration of the minimum distance of $LA_1 \cdots LA_2$ pairs and $LB_1 \cdots LB_2$ pairs to the $LA_1 \cdots LB_2$ distance (denoted as $(d_{AA}, d_{BB})_{min}/d_{AB}$ in Figure 1b). Based on the established criteria, the wurtzite structure shows great potential in obtaining natural SFLPs due to the lowest $(CN_{LA}, CN_{LB})_{max}$ value and the largest $(d_{AA}, d_{BB})_{min}/d_{AB}$ value (shown in Figure 1b). Therefore, the wurtzite crystal structure is under investigation in the following discussion.

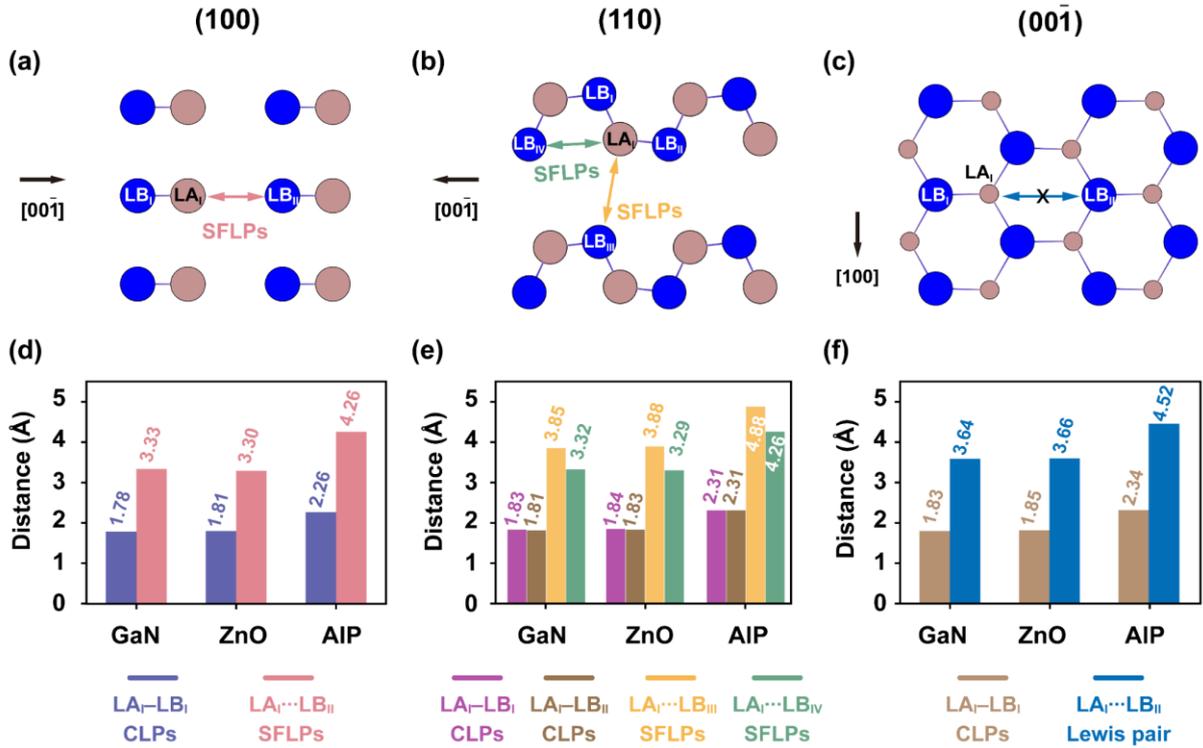

**Figure 2.** Identification of natural SFLPs on wurtzite crystal surfaces. Schematic images of (a) (100), (b) (110), and (c) (00$\bar{1}$) surfaces of wurtzite crystals. The distances of selected Lewis acid and Lewis base sites on (d) (100), (e) (110), and (f) (00$\bar{1}$) surfaces of wurtzite GaN, ZnO, and AlP. The small radii circles in (c) represent the cations in the second atomic layer.

The wurtzite-structured crystals, including GaN, ZnO, and AlP, are selected to explore the potential natural SFLPs on surfaces (see calculated lattice parameters in Table S2). Three low-index surfaces, i.e., (100), (110), and (00$\bar{1}$), are selected to analyze the possibility of SFLPs naturally existing. The schematic diagram of (100), (110), and (00$\bar{1}$) surfaces of wurtzite-crystal structures are shown in Figure 2a–c. On the (100) surface, two types of Lewis pairs are identified around $LA_I$ on the top atomic layer, namely $LA_I$–$LB_I$ with a shorter distance and $LA_I \cdots LB_{II}$ with a longer distance (Figure 2a). On the (100) surface of wurtzite GaN, ZnO, and AlP, the bonded $Ga_I$–$N_I$, $Zn_I$–$O_I$, and $Al_I$–$P_I$ pairs are identified as CLPs with shorter distances of 1.78 Å, 1.81 Å, and 2.26 Å, respectively. The unbonded $Ga_I \cdots N_{II}$, $Zn_I \cdots O_{II}$, and $Al_I \cdots P_{II}$ are identified as SFLPs due to no hindrance being found between the acid and base centers with longer acid-base distances of 3.33 Å, 3.30 Å, and 4.26 Å, respectively (Figure 2d and Figures S3–S8). As for the (110) surface, two kinds of unbonded Lewis pairs with longer distances, labeled as $LA_I \cdots LB_{III}$ and $LA_I \cdots LB_{IV}$, are confirmed as SFLPs because of no obstacle caused by the neighboring atoms, while the shorter $LA_I$–$LB_I$ and $LA_I$–$LB_{II}$ pairs are recognized as CLPs (Figure 2b). Specifically, the $Ga_I \cdots N_{III}$ and $Ga_I \cdots N_{IV}$ on GaN(110), $Zn_I \cdots O_{III}$ and $Zn_I \cdots O_{IV}$ on ZnO(110), and $Al_I \cdots P_{III}$ and $Al_I \cdots P_{IV}$ on AlP(110) are identified as FLPs with distances range from 3.29 Å to 4.88 Å (Figure 2e and Figures S3–S8). On the (00$\bar{1}$) surface, the $LA_I \cdots LB_{II}$ pairs, including $Ga_I \cdots N_{II}$ on GaN(00$\bar{1}$), $Zn_I \cdots O_{II}$ on ZnO(00$\bar{1}$), and $Al_I \cdots P_{II}$ on AlP(00$\bar{1}$), do not satisfy the concept of SFLPs due to the hindrance of the nearest-neighbor anions of $LA_I$, although its acid-base centers possess appropriate distances (Figure 2, c and f, and Figures S3–S8). Furthermore, to extend the concept of SFLPs on wurtzite crystal structures, the other twelve wurtzite crystals, including CdO, BaO, BeO, CdS, HgS, ZnS, MgS, InP, GaP, AlN, InN, and LaN, were also studied and found that on all the (100) and (110) surfaces of these crystals, SFLPs naturally existed with the acid-base center distances ranging from 2.86 to 5.27 Å (Figures S9–S12). Overall, after analyzing the atomic pairs on low-index wurtzite-structured surfaces, the natural SFLPs can be found on the (100) and (110) surfaces with the acid⋯base distances of 2.86–5.27 Å. Because the surface structures of wurtzite crystals are similar, in the following sections, only the GaN, ZnO, and AlP surfaces are taken as representatives to explore the characteristics of natural SFLPs in detail.

**Characteristics of Natural SFLPs on Wurtzite Crystal Surfaces**

Although natural SFLPs have been identified in geometry on the (100) and (110) surfaces of wurtzite crystals, the Lewis acidity/basicity is another characteristic to confirm the existence of SFLPs. As previously studied, the Lewis acidity and basicity can be evaluated by calculating the frontier orbitals, including the partial density of states (PDOS) and the band center, $\varepsilon$.[12] In general, the closer of the $\varepsilon$ to the Fermi level is, the stronger the Lewis acidity or the Lewis basicity will be. Figure S13 shows that the highest occupied molecular orbital (HOMO) of the GaN, ZnO, and AlP surfaces are mainly contributed by the N $2p_z$, O $2p_z$, and P $3p_z$ orbitals. The lowest unoccupied molecular orbital (LUMO) of the GaN, ZnO, and AlP surfaces are mainly contributed by the Ga $4p$, Zn $4s$, and Al $3p$ orbitals, respectively. Due to the similar coordination environment of surface atoms at the top atomic layer of wurtzite crystal surfaces, only one Lewis acid site and one Lewis base site need to be analyzed on each surface. On GaN(100), the acidity and basicity of the SFLPs ($Ga_I$ and $N_{II}$) can be quantitatively evaluated by $\varepsilon$ as 2.85 and –1.08 eV (Figure 3a), respectively. Similarly, as shown in Figure 3b, the $\varepsilon$ of $Ga_I$ $4p$ orbital and $N_{III}$ $2p_z$ orbital on GaN(110) is 3.31 and –1.16 eV, respectively. On the GaN(00$\bar{1}$) surface, the reactivity of the subsurface Ga atom is very weak, with a band center of 7.33 eV (Figure S14), suggesting the coordination-saturated Ga atoms on GaN(00$\bar{1}$) fail to act as Lewis acid sites. As shown in Figure 3e–f, the mean energy of valence electrons of O $2p_z$ orbital is measured as –1.13 eV on ZnO(100) and –1.09 eV on ZnO(110), respectively. However, the Zn $4s$ empty orbital possesses wide energy distributed ranging from 0.0 to 8.0 eV, causing a high mean energy of 4.29 eV on ZnO(100) and 4.80 eV on ZnO(110). As for wurtzite AlP, the energy of both the empty orbital of Al $3p$ and the lone pairs on the P $3p_z$ orbital are close to the Fermi level (Figure 3i and j). Quantitatively, the $\varepsilon$ of P $3p_z$ is –1.19 eV on the (100) and –1.31 eV on the (110), whereas the mean energy of Al $3p$ orbital is measured as 2.03 eV on AlP (100) and 2.36 eV on (110), respectively. Overall, besides the geometric structure, the appropriate Lewis acidity and basicity would enable SFLPs to be potential catalysts.

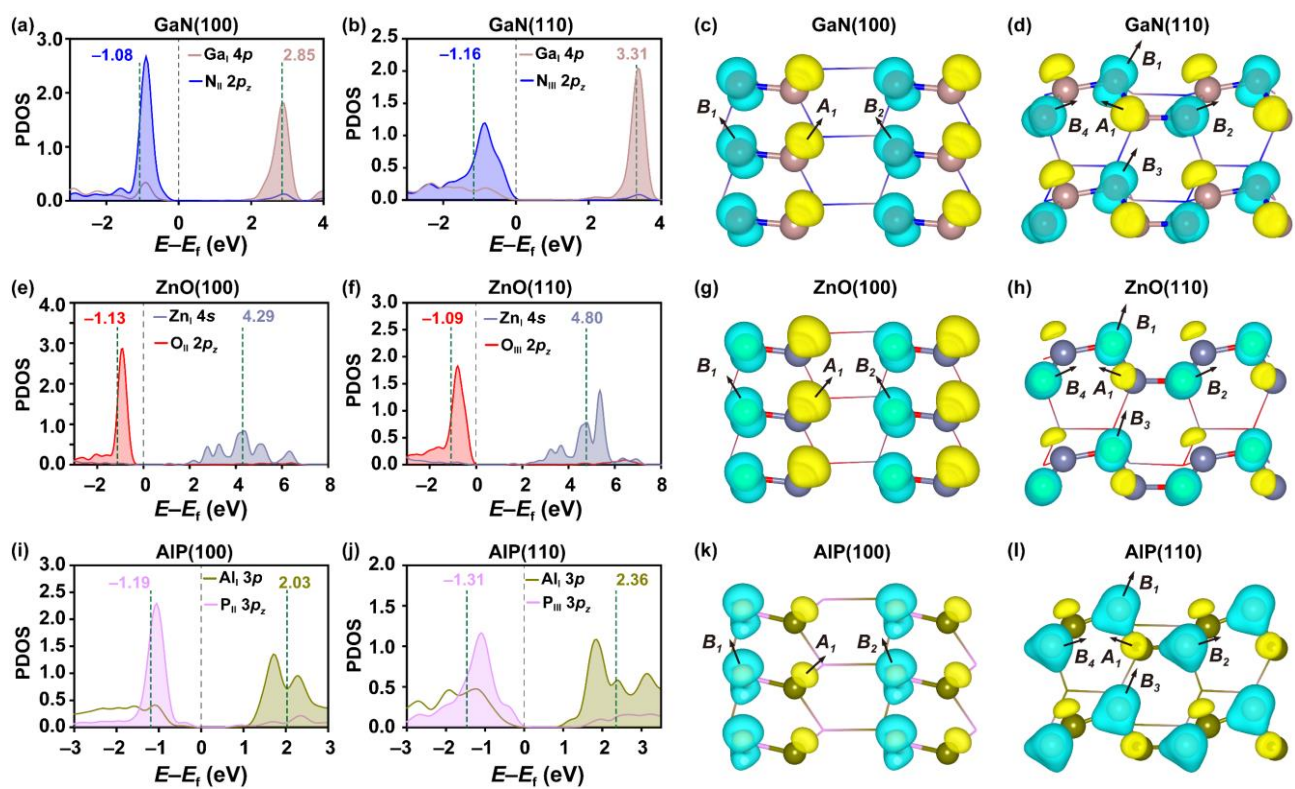

**Figure 3.** Lewis acidity, Lewis basicity, and frontier orbital orientation of natural SFLPs on wurtzite crystal surfaces. Partial density of states (PDOS) of selected atoms on (a) GaN(100), (b) GaN(110), (e) ZnO(100), (f) ZnO(110), (i) AlP(100), and (j) AlP(110) surfaces. Frontier orbitals maps of clean (c) GaN(100), (d) GaN(110), and (g) ZnO(100), (h) ZnO(110), (k) AlP(100), and (l) AlP(110) surfaces. The energy-weighted band centers are calculated using the equation: $\varepsilon = (\sum PDOS(E_i)\cdot E_i)/\sum PDOS(E_i)$, where PDOS($E_i$) is the PDOS of a given orbital in an energy interval ($E_i$, $E_i + \Delta E$) and $\Delta E$ is 0.05 eV. $E_i$ of N $2p_z$, O $2p_z$, and P $3p_z$ are in a range from −3.0 to 0.0 eV, while 0.0 to 4.0 eV for Ga $4p$, 0.0 to 8.0 eV for Zn $4s$, 0.0 to 3.0 eV for Al $3p$ on AlP(100), and 0.0 to 3.5 eV for Al $3p$ on AlP(110). The electron-density isosurfaces are plotted at 0.03 e/Bohr$^3$ for GaN surfaces and ZnO(100), 0.02 e/Bohr$^3$ for AlP surfaces, 0.04 and 0.05 e/Bohr$^3$ for LUMO and HOMO of ZnO(110), respectively. The brown, shadow blue, olive, blue, red, and pink balls represent Ga, Zn, Al, N, O, and P atoms, respectively. The yellow and cyan regions represent LUMO and HOMO, respectively. The arrows in the (c), (d), (g), (h), (k), and (l) pictures are used to indicate the direction of the vector.

Though individual Lewis acid/base sites possess identical acidity/basicity, when the Lewis acid and base sites cooperatively form Lewis pairs, the situation becomes different due to the diversity in frontier orbital orientations. In Figure 3c, the vector $A_1$ representing the orientation of the LUMO starts at $Ga_I$ and ends at the centroid of its LUMO. Similarly, the vector $B_2$ representing the orientation of the HOMO is formed by connecting $N_{II}$ and the centroid of the partial HOMO. The distance between centroids of HOMO and LUMO on $Ga_I \cdots N_{II}$ SFLPs is 2.74 Å (see details in the supplementary text and Tables S3−S4), which is shorter than the distance of the atomic pair (3.33 Å, Figure 2d), suggesting the LUMO and HOMO on $Ga_I \cdots N_{II}$ SFLPs tends to approach with each other. Similarly, the unbonded $Zn_I \cdots O_{II}$ and $Al_I \cdots P_{II}$ SFLPs on ZnO(100) and AlP(100) surfaces possess approached frontier orbitals, whereas the HOMO and LUMO of bonded $Zn_I-O_I$ and $Al_I-P_I$ pairs are keeping away from each other (Figure 3, g and k). Meanwhile, $Ga_I \cdots N_{III}$, $Ga_I \cdots N_{IV}$, $Zn_I \cdots O_{III}$, $Zn_I \cdots O_{IV}$, $Al_I \cdots P_{III}$, and $Al_I \cdots P_{IV}$ SFLPs on the (110) surfaces of wurtzite crystals also possess similar HOMO-LUMO spatial structures, as shown in Figure 3 d, h, and l. The above analysis demonstrates that the SFLPs on the (100) and (110) surfaces of wurtzite crystals possess approaching orientations of the frontier orbitals, which pave the way for the facile activation of small molecules.

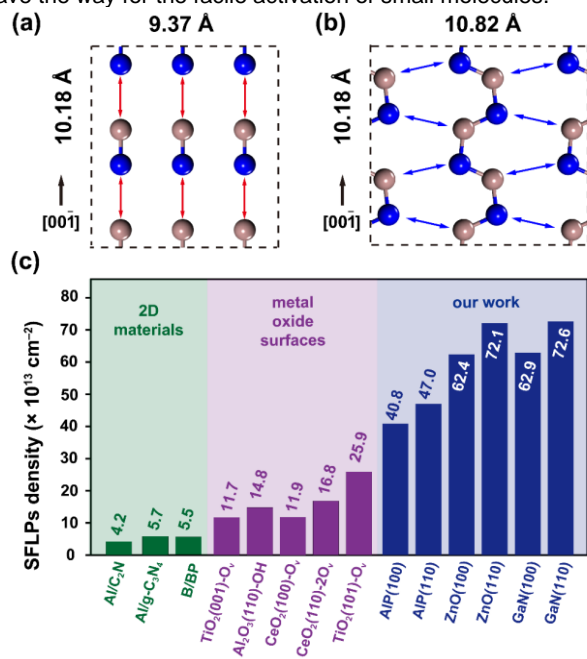

**Figure 4.** The density of SFLPs on wurtzite crystal surfaces. The distribution of SFLPs on (a) GaN(100) and (b) GaN(110) supercell. (c) The calculated surface density of SFLPs. The brown and blue balls represent Ga and N atoms, respectively.

The surface density of active sites is of great significance for the practical use of a catalyst. The surface density of SFLPs is calculated as the number of active sites per surface area. Six SFLPs are found on 3 × 2 supercell of GaN(100), ZnO(100), and AlP(100) surfaces, while eight SFLPs are found on GaN(100), ZnO(100), and AlP(100) 2 × 2 supercell (Figure 4a, 4b, and Figure S15). The surface density of SFLPs can reach up to $7.26 \times 10^{14}$ cm$^{-2}$. Interestingly, on the (100) and (110) surfaces, all the atoms on the top atomic layer serve as the frustrated Lewis acid/base sites, suggesting the unique atomic arrangements contribute to the high densities of the natural SFLPs. As shown in Figure 4c, previous SFLPs possess low densities ranging from $4.2 \times 10^{13}$ to $2.59 \times 10^{14}$ cm$^{-2}$,[6, 9b, 12a, 13] while our calculated density is nearly three times as much as the highest reported value. Taking the GaN surfaces as an example, the (100), (110), (001), (00$\bar{1}$), and (211) slabs are also taken to obtain a Wulff construction based on the surface energies.[14] The sum of the exposure ratio of {100} and {110} surfaces is 73.3% (see details in Figure S16), suggesting the two surfaces are predominated facets in practical crystals. Overall, the high density of SFLPs on wurtzite crystal surfaces contributed to all the surface atoms serving as sites of SFLPs, and the surfaces that can form natural SFLPs are identified to be most likely exposed.

The stability of a catalyst is another important indicator to evaluate whether it can be practically used. The bulk can maintain stability under high temperatures,[15] whereas the stability of active sites on the surfaces needs to be explored. Using ab initio molecular dynamics (AIMD) simulations under a high temperature of 800 K, the distance variations of Lewis pairs are calculated to evaluate the thermal stability. For GaN(100), the distance between $Ga_I$ and $N_{II}$ atoms ranges from 2.9 Å to 3.8 Å with a mean distance of 3.33 Å and a small standard deviation of $Ga_I \cdots N_{II}$ of 0.13 Å (Figure 5a, 5g and Table S5), demonstrating that the SFLP is stable at a high temperature of 800 K. The $LA_I \cdots LB_{II}$ distance ranges from 2.5 to 4.3 Å for the ZnO(100) and 3.3 to 4.9 Å for the AlP(100) with standard deviations of no more than 0.3 Å (Figure 5b, 5c, and Tables S7 and S9), suggesting that the SFLPs can be able to tolerate the high temperature of 800 K. Meanwhile, on the GaN(110), the $Ga_I \cdots N_{III}$ and $Ga_I \cdots N_{IV}$ SFLPs on the GaN(110) surface possess a moderate acid-base center distance from 3.2 to 4.4 Å and from 2.6 to 4.0 Å, respectively (Figure 5d). The standard deviation of $Ga_I \cdots N_{III}$ and $Ga_I \cdots N_{IV}$ is 0.17 and 0.18 Å (Figure 5h and Table S6), respectively. Similarly, the Zn⋯O on ZnO(110) and Al⋯P on AlP(110) also keep sufficient distance during the 10 ps AIMD simulation (Figure 5e–f). Quantitatively, the standard deviation of $LA_I \cdots LB_{III}$ and $LA_I \cdots LB_{IV}$ on the (110) surfaces is not more than 0.30 Å (Figure 5h, Tables S8 and S10), suggesting the SFLPs maintain good stability under the high temperature of 800 K. On the clean (100) and (110) surfaces of wurtzite GaN, ZnO, and AlP, the standard deviations of bonded CLPs are lower than 0.10 Å (Figure 5g and Figure S17), suggesting the limited variation of Ga–N, Zn–O, and Al–P bonds at the high temperature. The stable LA–LB dimer on the (100) surfaces and LA–LB chain on the (110) surfaces were other evidence for the stability of LA⋯LB SFLPs. Overall, the SFLPs on the clean surfaces possess excellent stability at high temperatures.

Furthermore, the stability of SFLPs in reactive atmospheres was also studied by AIMD simulations at 800 K. On the CO and H$_2$O covered GaN(100) surface, the bond length of $Ga_I$–$N_I$ and the Lewis acid-base center distance of $Ga_I \cdots N_{II}$ SFLPs are distributed in 1.6 to 2.3 Å and 2.6 to 3.9 Å (Figure S18), respectively. The standard deviation of $Ga_I \cdots N_{II}$ SFLPs under the CO and H$_2$O atmosphere are 0.17 Å and 0.19 Å, respectively, whereas only 0.10 Å and 0.08 Å for $Ga_I$–$N_I$ CLPs (Figure 5g). On the ZnO(100) and AlP(100) surfaces, as shown in Figures S20 and S22, the distance of $Zn_I \cdots O_{II}$ and $Al_I \cdots P_{II}$ ranges from 2.4 to 4.1 Å and from 3.4 to 4.8 Å, respectively, indicating that the SFLPs on the surfaces can resist the high temperature and reaction atmospheres. Similarly, the stability of SFLP sites on GaN(110), ZnO(110), and AlP(110) surfaces is also proved by small variation and standard deviations of LA–LB and LA⋯LB Lewis pairs (Figure 5h and Figures S17, S19, S21, and S23). In short, the SFLPs on (100) and (110) surfaces of wurtzite crystals can remain stable under high temperatures and the adsorption of reactive molecules such as CO and $H_2O$.

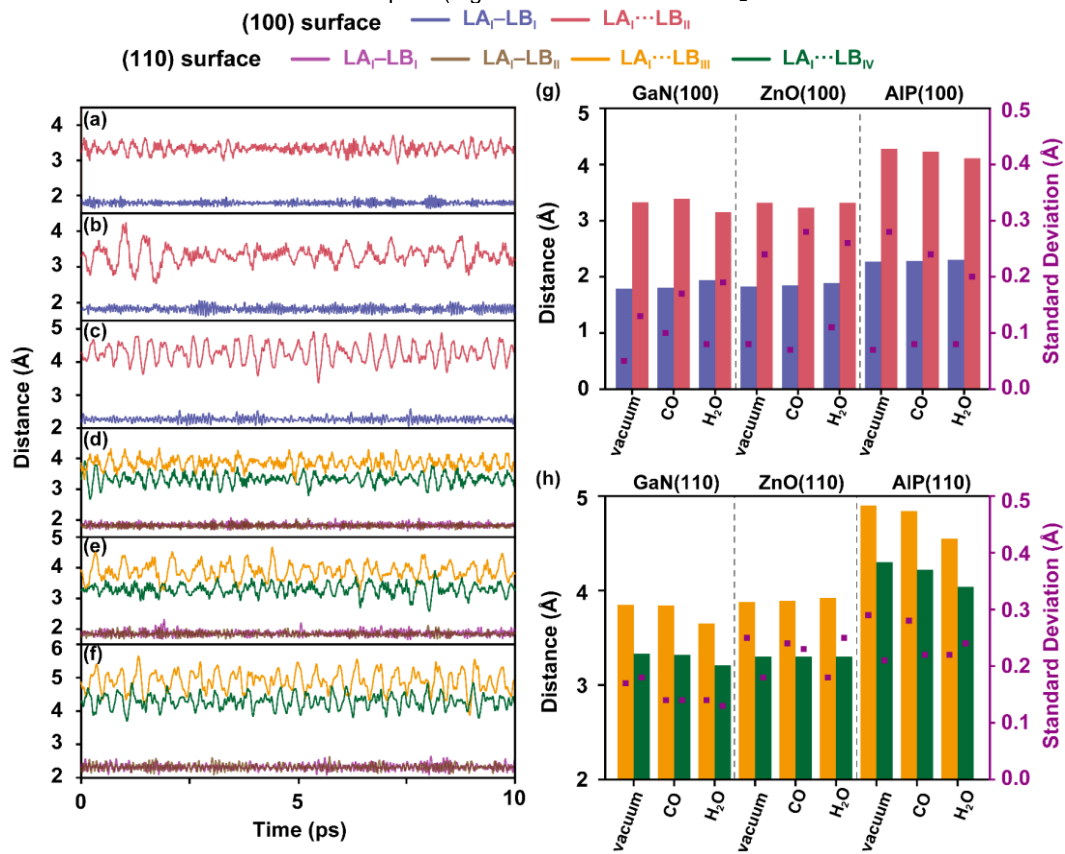

**Figure 5.** Stability analysis of natural SFLPs on wurtzite crystal surfaces. Variation of the distance between selected acid and base sites on clean (a) GaN(100), (b) ZnO(100), (c) AlP(100), (d) GaN(110), (e) ZnO(110), and (f) AlP(110) upon AIMD simulations at 800 K. The mean distance and standard deviation of selected Lewis pairs on the (g) (100) surfaces and (h) (110) surfaces of wurtzite crystals during the AIMD simulations at 800 K with different reaction atmospheres.

## Performance of natural SFLPs on small molecules activation

The activation of small molecules (including $H_2$, $CH_4$, $NH_3$, $H_2S$, and $PH_3$) at SFLPs on (100) surfaces of GaN, ZnO, and AlP was performed. The activation energy ($E_a$) of small molecules dissociation was selected as the criteria for evaluation. Pt was selected as a reference catalyst due to its excellent catalytic performance for the activation of a series of small molecules.[16] For hydrogen dissociation, the SFLPs on GaN and ZnO exhibit comparable performance with Pt (Figure 6), which is in good agreement with our previous results.[17] While on the SFLPs of AlP, the $H_2$ dissociation presents a relatively high barrier of 0.67 eV. The related potential energy profiles and optimized intermediates and transition states (TS) are given in Figures S24–S31. As for the methane dissociation, the activation energy is 0.62 eV for SFLPs on GaN(100) and 0.82 eV for SFLPs on ZnO(100), suggesting those SFLPs perform excellent C–H bond cleave ability as good as Pt(111) with an activation energy of 0.69 eV. Unlike hydrogen and methane, the polar $NH_3$ molecule chemically adsorbs on the LA with large adsorption energy ranging from –1.50 eV to –1.58 eV (Figure S24). The activation energy is only 0.16 eV and 1.05 eV on SFLPs of GaN and AlP, respectively, significantly lower than that on Pt(111) with an $E_a$ of 1.37 eV. In addition, the *H, *$NH_2$ intermediate cannot form on SFLPs of ZnO(100), which may be caused by the unfavorable thermodynamics of the intermediate. The dissociation of $H_2S$ can easily occur on all the SFLPs of GaN, ZnO, and AlP with activation barriers of lower than 0.1 eV, even lower than that (0.12 eV) on Pt(111). As for $PH_3$, the activation barrier is calculated to be 0.10 eV for SFLPs on GaN and 0.28 eV for SFLPs on ZnO, which is comparable to or better than that on Pt(111). On AlP(100), the P–H bond of $PH_3$ dissociation needs to overcome an energy barrier of 0.51 eV, suggesting the facile activation process can also be achieved at SFLPs of AlP. In summary, SFLPs on GaN can activate all five molecules (including $H_2$, $CH_4$, $NH_3$, $H_2S$, and $PH_3$) with excellent catalytic performance, and the SFLPs on ZnO are remarkable for $H_2$, $CH_4$, $H_2S$, and $PH_3$ activation, while AlP can play an effective role in $NH_3$, $H_2S$, and $PH_3$ activation. Based on the different performances of the different SFLPs, we can find that the reactivity of the active sites depends not only on the geometry structure but also on the chemical composition as well as the matching degree with activation substrates.

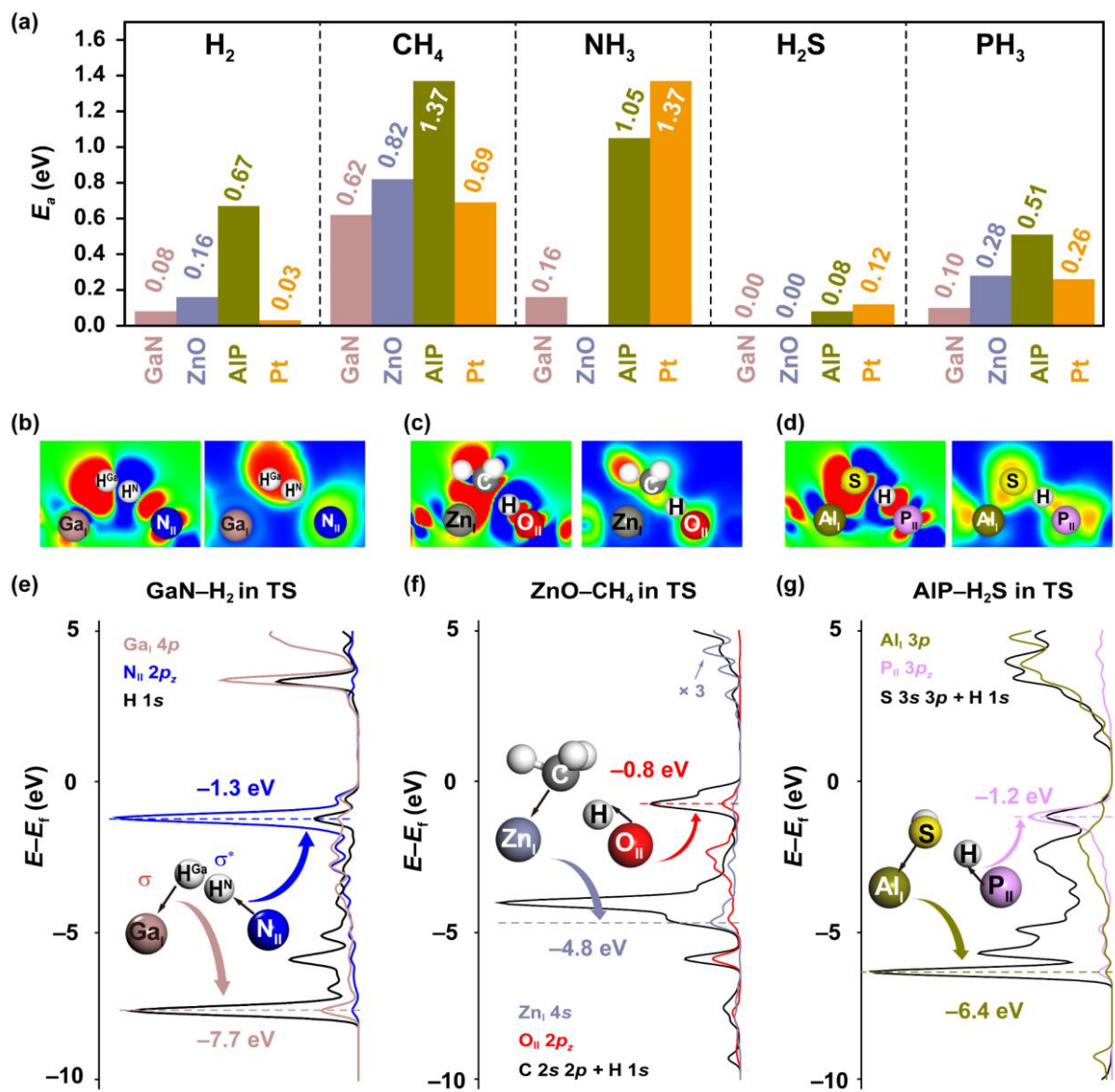

**Figure 6.** Small molecule activation on wurtzite crystal surfaces. (a) The activation energy of hydrogen, methane, ammonia, hydrogen sulfide, and phosphine dissociation on SFLPs of wurtzite crystal surfaces and Pt(111). Electron density difference ($\Delta\rho = \rho(TS) - \rho(surface^{\#}) - \rho(molecule^{\#})$) maps (left) and their corresponding electron localization function map (right) for (b) hydrogen dissociation on SFLPs of GaN(100), (c) methane dissociation on SFLPs of ZnO(100), and (d) hydrogen sulfide dissociation on SFLPs of AlP(100). PDOS analysis in TS of (e) hydrogen, (f) methane, and (g) hydrogen sulfide dissociation on wurtzite crystal surfaces. The charge density difference and electron localization function map were calculated with a range from –0.002 to 0.002 e/Bohr$^3$, and 0.0 to 1.0, respectively. In charge density difference map, the red and blue areas represent accumulation and depletion of charges, respectively. The atomic positions of surface$^{\#}$ and molecule$^{\#}$ are identical to those in the TS. The Blue, green, and red colors correspond to the ELF values of 0.0, 0.5, and 1.0, respectively. All the hydrogen atoms of methane and hydrogen sulfide are considered in PDOS analysis.

The TS of hydrogen dissociation on GaN(100), methane dissociation on ZnO(100), and H$_2$S dissociation on AlP(100) are selected to comprehensively analyze the catalytic principle of the SFLPs. The Lewis acid–small molecule–Lewis base fragments in TS are analyzed to understand the orbital overlap between small molecules and SFLPs. A previous study reported that the optimal orbital overlap of acid–H$_2$ is a side-on mode, while an end-on mode is applied for the base–H$_2$ interaction (Figure S32), which contributes to the electron transfer from the Lewis base to σ*(H$_2$) and σ(H$_2$) to the Lewis acid.[18] The activation of other σ bonds, such as C–H and S–H bonds, is followed by a similar interaction between small molecules and active sites. As for the TS of hydrogen dissociation on SFLPs of GaN(100), the angle H$^{Ga}$–H$^N$–N$_{II}$ and H$^N$–H$^{Ga}$–Ga$_I$ are 165.9° and 100.8° (Table S11), respectively, suggesting the side-on acid–H$_2$ mode and end-on base–H$_2$ mode are simultaneously achieved on SFLPs (Figure 7e). Meanwhile, during the methane dissociation, the angles C–H–O$_{II}$ and H–C–Zn$_I$ on SFLP of ZnO(100) are 174.9° and 73.7° (Figure 7f and Table S12), respectively, and the angles S–H–P$_{II}$ and H–S–Al$_I$ are measured as 161.7° and 95.1° in the H$_2$S activation (Figure 7g and Table S13), respectively. The results above indicate that the optimal geometric configuration during small molecule activation is also achieved on SFLPs of ZnO(100) and AlP(100).

To understand the interaction between SFLPs and small molecules, the electronic structure of small molecules activation on SFLPs of wurtzite GaN, ZnO, and AlP, including charge density difference, electron localization function (ELF), and PDOS analysis, were calculated. In Figure 7b, the charge density difference maps show that the electron density of $H_2$ bonding state decreases, whereas the electron density of $H_2$ antibonding state increases. Moreover, the ELF map shows that electrons between $Ga_I$ the $H^{Ga}$ are highly delocalized, whereas localized regions exist around the $N_{II}\cdots H^N$ (Figure 7b), suggesting Coulombic attraction and covalent bonding mainly exist in $Ga_I\cdots H^{Ga}$ and $N_{II}\cdots H^N$, respectively. As shown in Figure 7e, the energy levels of the $N_{II}$ $2p_z$ orbital and $H_2$ $\sigma^*$ orbital in TS are in good consistence (at $E–E_f = –1.3$ eV), and a p-σ orbital overlap is formed at a lower energy level (at $E–E_f = –7.7$ eV), which confirms the proposed electrons transfer pathways between SFLPs and $H_2$ (Figure S32). As a comparison, on CLPs, the p-σ* orbital overlap is formed at a higher energy level ($E–E_f = –0.9$ eV, as shown in Figure S33), indicating the advantages of SFLPs. As for methane activation on SFLPs of ZnO(100), charge density difference maps show the electrons accumulate in the methyl group, whereas the electrons deplete around the H atom (Figure 7c). In Figure 7f, PDOS shows the C–H antibonding orbital accepts electrons from Lewis base (at $E–E_f = –0.8$ eV), while the C–H bonding orbital donates electrons to Lewis acid (at $E–E_f = –4.8$ eV). Figure 7d shows the electrons accumulate between $P_{II}$ and H, suggesting that the H atom strongly binds with the surface. An orbital overlap can be found at $E–E_f = –1.2$ eV for hydrogen sulfide activation on SFLPs of AlP, representing the electrons transfer from the Lewis base ($P_{II}$) to the S–H antibonding orbital. Meanwhile, a few electrons from hydrogen sulfide donated to the Lewis acid ($Al_I$), forming a peak at $E–E_f = –6.4$ eV (Figure 7g). Furthermore, the electron transfer process was also explored to understand other small molecule and SFLP site interactions (Figures S34–S35). Overall, the approached orbital orientation of SFLPs contributes to the optimal orbital overlap between SFLPs and small molecules.

## Conclusion

Wurtzite-structured crystals such as GaN, ZnO, and AlP are found as ideal materials for developing natural surface frustrated Lewis pairs with high surface density and outstanding stability. It is found that all the surface atoms on wurtzite-structured (100) and (110) facets can serve as frustrated Lewis acid/base sites, contributing to the high surface density of SFLPs up to $7.26 \times 10^{14}$ cm$^{-2}$. The distance variations during AIMD simulations demonstrate that the natural SFLPs on GaN, ZnO, and AlP surfaces possess remarkable stability under high temperatures and reactive atmospheres of CO and $H_2O$. The extraordinary performance of the SFLPs on GaN, ZnO, and AlP for the activation of a series of small molecules, including $H_2$, $CH_4$, $NH_3$, $H_2S$, and $PH_3$, was demonstrated. Electronic structure analysis indicates that the approaching orbital orientation of SFLPs contributes to the optimal SFLPs and small molecule orbital overlap. To sum up, this work provides a simple and efficient method to obtain dense and stable surface FLPs on a common type of solid material, i.e., wurtzite-structured crystals, and unravels the novel feature of SFLPs in activating small molecules.

## Supporting Information

The authors have cited additional references within the Supporting Information.[19-30]

## Acknowledgements

This work is supported by the National Natural Science Foundation of China (22078257 and U23A20112), the National Key R&D Program of China (2023YFA1506300), and the Joint Fund of the Yulin University and the Dalian National Laboratory for Clean Energy (Grant No. YLU-DNL Fund 2022001). C.-R. C. also acknowledges the support from the Young Talent Fund of Shaanxi Province. The calculations were performed by using the HPC Platform at Xi'an Jiaotong University.

**Keywords:** Surface frustrated Lewis pairs • Natural • Wurtzite • Small molecules

Supporting Information

# Finding Natural, Dense, and Stable Frustrated Lewis Pairs on Wurtzite Crystal Surfaces

Xi-Yang Yu,[a] Zheng-Qing Huang,[a] Tao Ban,[a] Yun-Hua Xu,[b] Zhong-Wen Liu,*[c] and Chun-Ran Chang*[a,b]

| | |
|---|---|
| [a] | X.-Y, Yu, Dr. Z.-Q., Huang, T. Ban, Prof. Dr. C.-R., Chang<br>Shaanxi Key Laboratory of Energy Chemical Process Intensification<br>School of Chemical Engineering and Technology, Xi'an Jiaotong University<br>Xi'an 710049 (China)<br>E-mail: changcr@mail.xjtu.edu.cn |
| [b] | Prof. Dr. Y.-H., Xu, Prof. Dr. C.-R., Chang<br>Shaanxi Key Laboratory of Low Metamorphic Coal Clean Utilization<br>School of Chemistry and Chemical Engineering, Yulin University<br>Yulin 719000 (China) |
| [c] | Prof. Dr. Z.-W., Liu<br>Key Laboratory of Syngas Conversion of Shaanxi Province<br>School of Chemistry and Chemical Engineering, Shaanxi Normal University<br>Xi'an 710119 (China)<br>E-mail: zwliu@snnu.edu.cn |

## Computational Details

### Electronic structure methods

All the spin-polarized DFT calculations were performed by using the Vienna ab initio simulation package (VASP).[19] The Perdew-Burke-Ernzerhof (PBE) functional with the generalized gradient approximation (GGA) was used to treat the exchange-correlation potential.[20] The kinetic energy cutoff of 400 eV was used. The projector augmented wave (PAW) pseudopotential was adopted to treat the core electrons,[21] and the $3d$ electrons of Ga atoms, $4d$ electrons of In atoms, $5s$ electrons of Ba atoms, $1s$ electrons of Be atoms, and $2p$ electrons of Mg atoms are treated as valence electrons. The Coulomb repulsion and exchange interaction of $3d$ electrons of Ga of wurtzite GaN, Zn of wurtzite ZnO were treated using the DFT + $U$ method with effective $U$ values of 3.9 eV and 7.5 eV, respectively.[22] Zero damping DFT + D3 correction method was utilized to account for *van der Waals* interaction of adsorbates and surfaces.[23] A 9 × 9 × 9 Monkhorst-Pack mesh *k*-point was used to optimize bulk structures of wurtzite crystal.[24] The calculated results are shown in Table S2, which was in good agreement with the previous works.[25] The convergence criteria for force on ions and energy were set as 0.02 eV/Å and $10^{-5}$ eV, respectively.

The adsorption energy, termed $E_{ads}$, was calculated by using the following formula equation, $E_{ads} = E_{adsorbate+molecule} − E_{surface} − E_{adsorbate}$. The $E_{adsorbate+molecule}$ is the total energy of the adsorbate covered on the surfaces, where the $E_{surface}$ and $E_{adsorbate}$ represent the energy of the clean surface and adsorbates, respectively. The reaction energy (Δ$E$) was calculated by the energy difference between the final state and its corresponding initial state. The activation energy ($E_a$) was determined as the energy difference between the transition state and its corresponding initial state. The nudged elastic band combined with the minimum-mode following the dimer method was adopted for transition state calculation.[26] The vibration analysis verified all the transition states. The LOBSTER package was used to calculate the projected crystal orbital Hamilton population curves and projected density of states.[27] The pbeVASPFit2015 basis set was adopted for the projection wave function.[27d]

### Surface models

The vacuum region was set as 15 Å in the perpendicular direction for both 3 × 2 supercell with six layers for (100) surfaces of wurtzite crystals and 2 × 2 supercell with five layers for (110) surfaces of wurtzite crystals. While the vacuum region with 20 Å in the perpendicular direction was set for 3 × 3 supercell with eight layers of polar GaN(00$\bar{1}$), ZnO(00$\bar{1}$), and AlP(00$\bar{1}$) surfaces. The Brillouin zone was sampled with a 2 × 2 × 1 *k*-point grid for (100), (110), and (00$\bar{1}$) surfaces. As for Pt(111) surface, a 15 Å vacuum region was set in the perpendicular direction for 4 × 4 supercell with four layers. A 3 × 3 × 1 *k*-point grid was used for Pt(111) surface. The bottom two layers of the slab were fixed, while the atoms on the other layers with adsorbates were allowed to relax.

The formula of surface energy can be written as follows:

$$\gamma = \frac{1}{A}[E_{sur\_relax} - NE_{bulk}] - \frac{1}{2A}[E_{sur\_unrelax} - NE_{bulk}]$$

where $E_{sur\_relax}$ can represent the total energy of GaN surfaces, $E_{sur\_unrelax}$ is the total energy of the unrelaxed surface, $E_{bulk}$ is the total energy of the GaN bulk structure, N is the number of GaN units in the cell, A is the area of the GaN surface considered.

The chemical potential of $N_2$ is directly defined by the temperature and pressure of nitrogen. By treating the $N_2$ as an ideal gas, the formula of chemical potential can be calculated as follows:

$$\mu_{N_2} = \mu_{N_2}^o(T, P^o) + kT \ln(P_{N_2}/P_{N_2}^o)$$

where superscript o represents the standard pressure. The chemical potential of gaseous $N_2$ ($\mu_{N_2}$) and atomic N ($\mu_N$) in the GaN bulk is calculated by:

$$\mu_N = \frac{1}{2}\mu_{N_2}$$

In our calculations, the temperature is set as 0 K, and the pressure is 0 bar. Therefore, the chemical potential of N can be considered as

$$\mu_N = \frac{1}{2}E_{N_2}$$

Based on the surface energies, the thermodynamic equilibrium morphology of a freestanding particle can be calculated according to the Gibbs–Wulff theorem.[14] The Wulff construction is taken as follows: beginning from the center point, a GaN surface that is normal to the vector ***hkl*** is drawn at a distance of $d_{hkl} = C × \gamma_{hkl}$. C and γ$_{hkl}$ are a constant value and the surface energy of unit area that normal to vector **hkl**, respectively. After this process is repeated for every Miller-index surface, the equilibrium morphology of the GaN crystal is given.

**Ab initio molecular dynamics simulations**

Ab initio molecular dynamics (AIMD) simulations were calculated with the VASP package to evaluate the stability of the GaN (100) and (110) surfaces at a high temperature of 800 K. The 1 × 1 × 1 k-points was used for AIMD simulations. The coverages of the CO and $H_2O$ were set as 1/2ML. All the AIMD simulations were carried out for 10 ps with a time step of 0.5 fs by using Nosé–Hoover thermostats and canonical (*NVT*) ensemble.[28]

**Radial distribution function calculations**

The radial distribution function (RDF), *g(r)*, is the probability that finding the center of a particle varies as a function of the distance from the given center.[29] The intensity *n(r)* is calculated by counting the number of Lewis acid and Lewis base centers between the given ranges of separation. The general expression of *n(r)* is: $n(r) = \int_{r}^{r+\Delta r} g(r)\, 4\pi r^2 dr$, where $\Delta r$ is set as 0.1 Å. The CuO crystals structure were referred from Inorganic Crystal Structure Database, whereas the rest thirteen structures for RDF analysis were referred from Materials Project database.[30] The details of crystal structures are shown in Table S1.

## Results and Discussion

**Calculation details of centroids of frontier orbital**

The charge density file is obtained from the PARCHG file through DFT calculation. For GaN(100) surface, the energy range of HOMO and LUMO is –1.2 to –0.6 eV and 2.0 to 3.5 eV, respectively. The LUMO is completely taken into calculation. As for HOMO, the dumbbell-shaped orbital around N is partially taken into calculation. Only the charge density above the base sites is considered because those electrons are mainly involved in small molecule activation. The formula for calculating the centroid coordinates is:

$$i = \frac{\int \rho \cdot r\, d\rho}{\int \rho\, d\rho}$$

$\rho$ represents the charge density, and r means the corresponding coordinate. The coordinates of Lewis sites and the calculated centroids of the frontier orbitals are listed in Table S3.

**Vector Direction Study on GaN(100) Surface**

The orbital orientation of the HOMO and LUMO are quantitatively investigated by calculating the vector's direction. ***l$_1$*** is the normal vector of the plane consisting of ***A$_1$*** and ***Ga$_l$N$_{ll}$***, and ***l$_2$*** is the normal vector of the plane consisting of ***B$_1$*** and ***Ga$_l$N$_{ll}$***. If the vectors ***l$_1$*** and ***l$_2$*** possess the same direction, ***A$_1$*** and ***B$_2$*** are coplanar. Otherwise, it proves that the ***A$_1$*** and ***B$_2$*** are not coplanar.

According to the coordinate of Lewis sites and the calculated centroids of the frontier orbitals (Table S3), the dimensionless vectors ***A$_1$***, ***B$_2$***, and ***Ga$_l$N$_{ll}$*** are calculated as (0.000, 0.039, 0.045), (0.000, 0.019, 0.023), and (0.000, −0.326, 0.010), respectively. The normal vector ***l$_1$*** is calculated as ***l$_1$*** = ***A$_1$*** × ***Ga$_l$N$_{ll}$*** = $\begin{vmatrix} i & j & k \\ 0.000 & 0.039 & 0.045 \\ 0.000 & -0.326 & 0.010 \end{vmatrix}$ = (0.01506, 0.00000, 0.00000), and ***l$_2$*** is calculated as ***l$_2$*** = ***B$_2$*** × ***Ga$_l$N$_{ll}$*** = $\begin{vmatrix} i & j & k \\ 0 & 0.019 & 0.023 \\ 0 & -0.326 & 0.010 \end{vmatrix}$ = (0.00769, 0.00000, 0.00000), while ***i*** = (1, 0, 0), ***j*** = (0, 1, 0), and ***k*** = (0, 0, 1). Due to $\frac{l_1 \cdot l_2}{|l_1| \times |l_2|}$ = 1, the vectors ***l$_1$*** and ***l$_2$*** possess the same direction. The above calculation shows that vectors ***A$_1$*** and ***B$_2$*** on GaN(100) are coplanar.

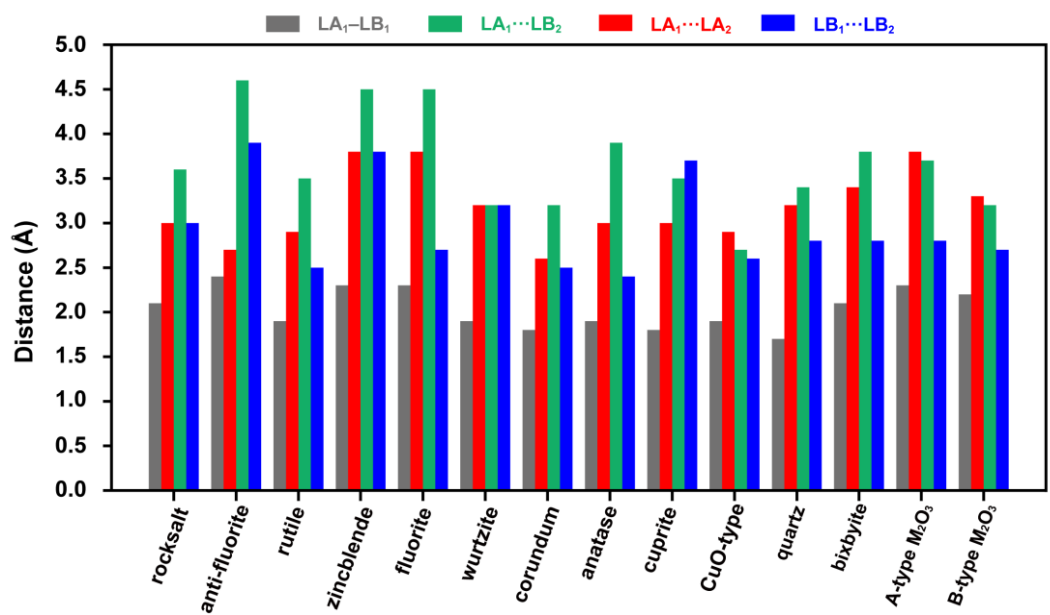

**Figure S1.** Radial distribution function of various bulk crystal structures. The distances between atom centers of fluorite, rocksalt, rutile, anti-fluorite, zincblende, wurtzite, corundum, anatase, cuprite, CuO-type, quartz, bixbyite, A-type $M_2O_3$ (M represent metal) and B-type $M_2O_3$ crystal structures.

**Relationship between the ratio of $d_{AA}$($d_{BB}$) to $d_{AB}$ and potential hindrance**

As shown in Figure S2a–b, a triangle with edge length $d_{AB}$, $d_{AA}$, $r_{AB}$ can represent the distance between LA$_1$ and its next-nearest-neighbor anion (LB$_2$), the distance between LA$_1$ and its nearest-neighbor cation (LA$_2$), and the bond length of LA$_1$–LB$_1$. Θ is the angle between $d_{AA}$ and $r_{AB}$. $l$ starts from the LA$_2$ point and is perpendicular to $d_{AB}$. The larger the value of $l$ is, the smaller the potential hindrance is. The formula for $l$ can be calculated as:

$$l = \frac{r_{AB} d_{AA}}{d_{AB}} \sin\theta$$

$$l^2 = \frac{r_{AB}^2 d_{AA}^2}{d_{AB}^2} (1 - \cos^2\theta)$$

where the $\cos\theta$ can be calculated as follows:

$$\cos\theta = \frac{d_{AA}^2 + r_{AB}^2 - d_{AB}^2}{2 r_{AB} d_{AA}}$$

Therefore, the equation for $l$ can be written as follows:

$$\frac{4l^2}{d_{AB}^2} = \frac{4 r_{AB}^2 d_{AA}^2}{d_{AB}^4} - \frac{(d_{AA}^2 + r_{AB}^2 - d_{AB}^2)^2}{d_{AB}^4}$$

The ratio of $r_{AB}$ to $d_{AB}$ is set as the constant a, the independent variable $x$ and dependent variable $y$ are defined as follows:

$$x = \frac{d_{AA}}{d_{AB}}; a = \frac{r_{AB}}{d_{AB}}; y = f(x) = \frac{4l^2}{d_{AB}^2}$$

The expression for $y$ and the derivative of $y$ can be calculated as follows:

$$y = f(x) = -x^4 + 2a^2 x^2 + 2x^2 - a^4 + 2a^2 - 1$$

$$\frac{dy}{dx} = 4x(1 + a^2 - x^2)$$

The value of y will increase as the value of x increases if the value of $(1 + a^2 - x^2)$ is positive (x > 0). Based on the data shown in Figure S1, the value of a ranges from 0.49 to 0.70, and the $d_{AA}/d_{AB}$ distributed from 0.58 to 1.08, suggesting the value of $(1 + a^2 - x^2)$ is always positive. Therefore, the potential hindrance will decrease as $d_{AA}/d_{AB}$ increases.

As for the pontential hindrance caused by the LB$_1$ (Figure S2c–d), the expression can be written as follow:

$$\frac{4l^2}{d_{AB}^2} = \frac{4 r_{AB}^2 d_{BB}^2}{d_{AB}^4} - \frac{(d_{BB}^2 + r_{AB}^2 - d_{AB}^2)^2}{d_{AB}^4}$$

The $d_{AA}/d_{AB}$ ranges from 0.60 to 1.06. Similarly, the larger the $d_{BB}/d_{AB}$ is, the smaller the potential hindrance will be.

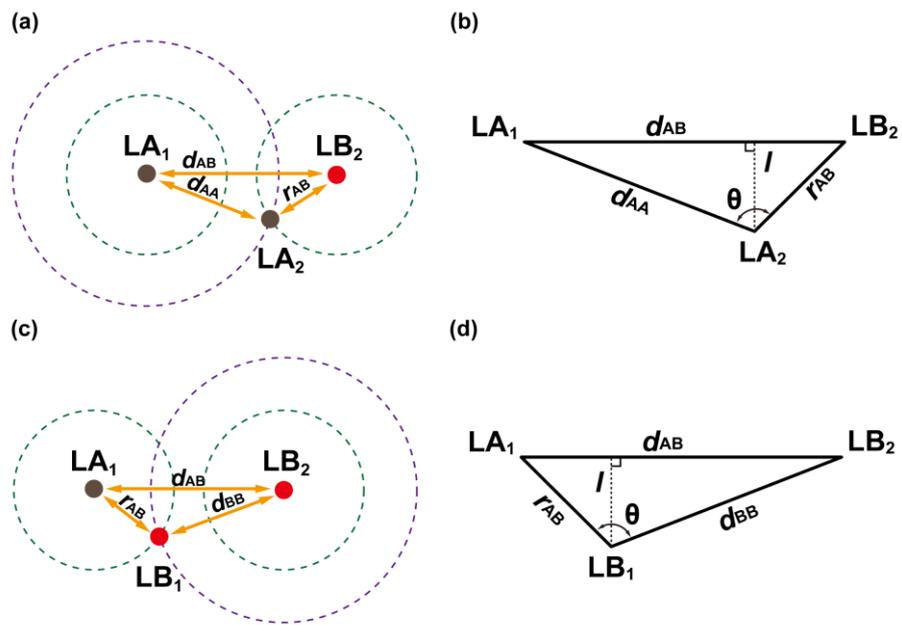

**Figure S2.** Schematic image of potential hindrance cause by (a–b) $LA_2$ and (c–d) $LB_1$ atom.

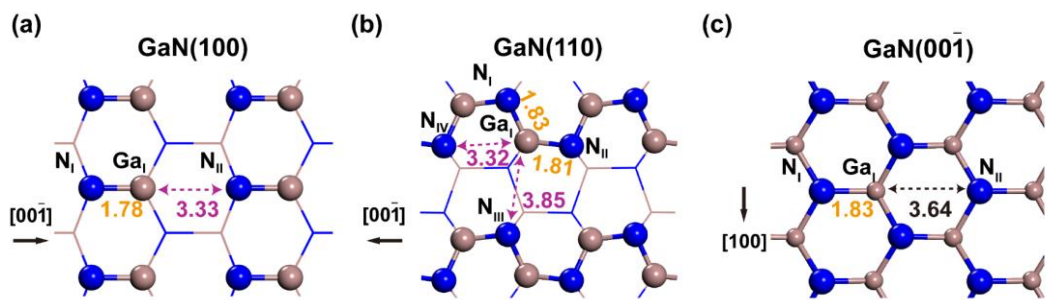

**Figure S3.** Optimized structures of (a) GaN(100), (b) GaN(110), and (c) GaN(00$\bar{1}$). The distances between Lewis acid and Lewis base labeled inside the picture are in Å. The natural SFLPs are labeled by dashed arrows. The distances between Lewis acid and Lewis base are in Å.

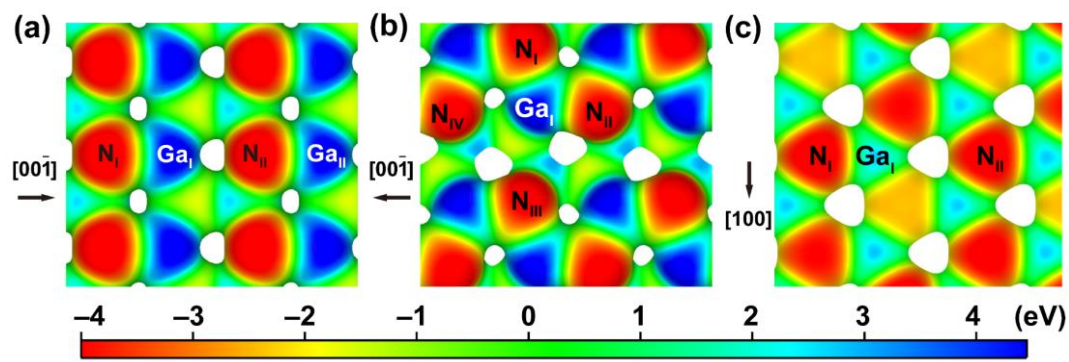

**Figure S4.** Charge density distribution of GaN surfaces. Electron-density isosurface of (a) GaN(100), (b) GaN(110), and (c) GaN(00$\bar{1}$) surfaces. The electron-density isosurfaces are plotted at 0.02 e Bohr$^{-3}$. The color bar represents the electrostatic potential scale.

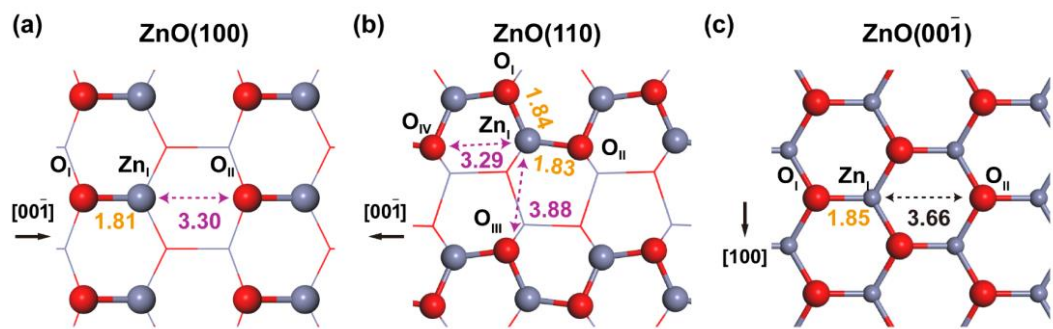

**Figure S5.** Optimized structures of (a) ZnO(100), (b) ZnO(110), and (c) ZnO(00$\bar{1}$). The distances between Lewis acid and Lewis base labeled inside the picture are in Å. The natural SFLPs are labeled by dashed arrows. The distances between Lewis acid and Lewis base are in Å.

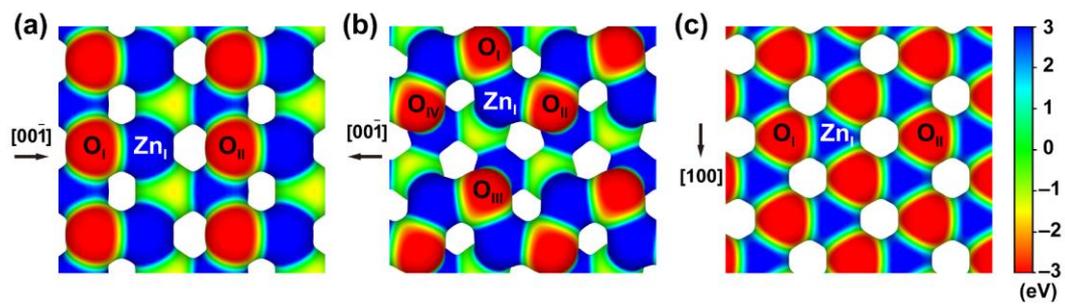

**Figure S6.** Charge density distribution of ZnO surfaces. Electron-density isosurface of (a) ZnO(100), (b) ZnO(110), and (c) ZnO(00$\bar{1}$) surfaces. The electron-density isosurfaces are plotted at 0.03 e Bohr$^{-3}$. The color bar represents the electrostatic potential scale.

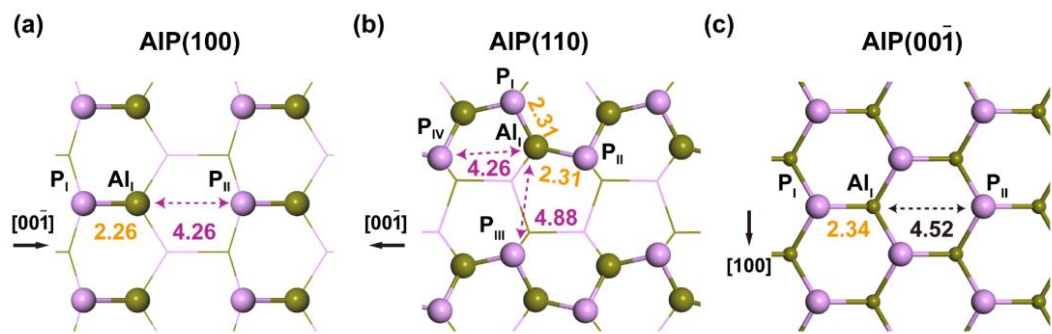

**Figure S7.** Optimized structures of (a) AlP(100), (b) AlP(110), and (c) AlP(00$\bar{1}$). The distances between Lewis acid and Lewis base labeled inside the picture are in Å. The natural SFLPs are labeled by dashed arrows. The distances between Lewis acid and Lewis base are in Å.

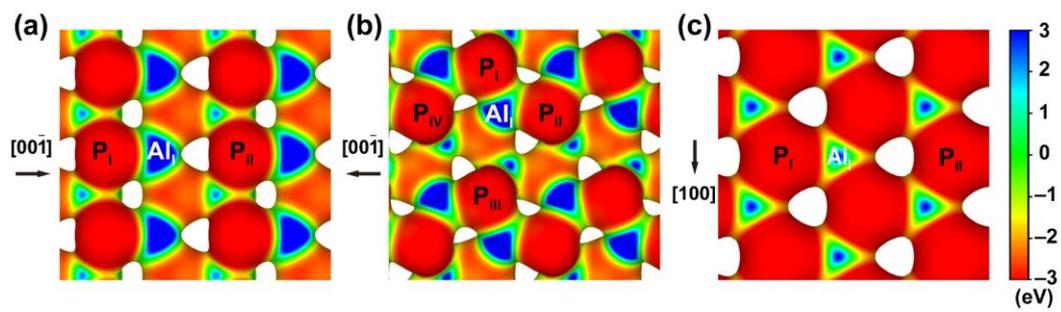

**Figure S8.** Charge density distribution of AlP surfaces. Electron-density isosurface of (a) AlP(100), (b) AlP(110), and (c) AlP(00$\bar{1}$) surfaces. The electron-density isosurfaces are plotted at 0.014 e Bohr$^{-3}$. The color bar represents the electrostatic potential scale.

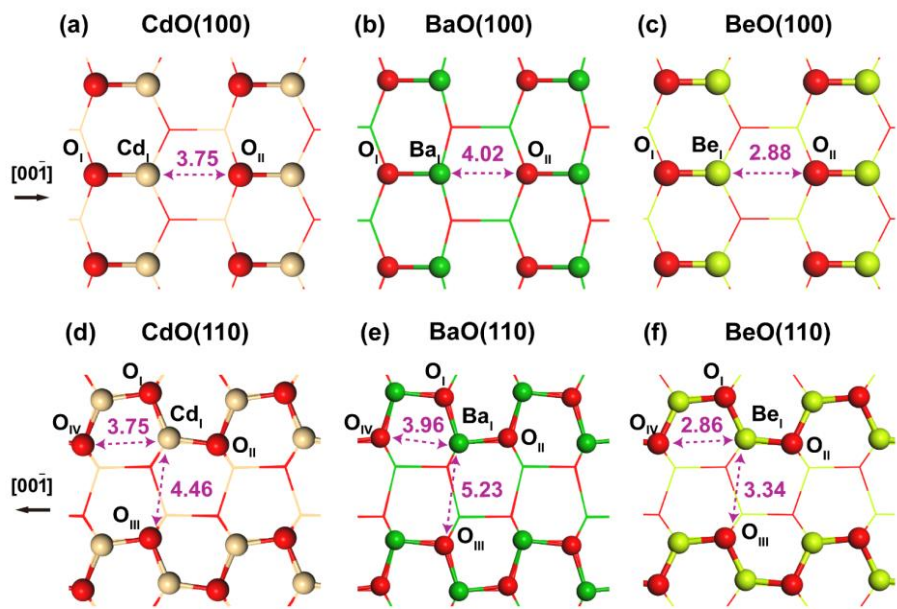

**Figure S9.** Optimized structures of (a) CdO(100), (b) BaO(100), (c) BeO(100), (d) CdO(110), (e) BaO(110) and (f) BeO(110). The natural SFLPs are labeled by dashed arrows. The distances between Lewis acid and Lewis base are in Å.

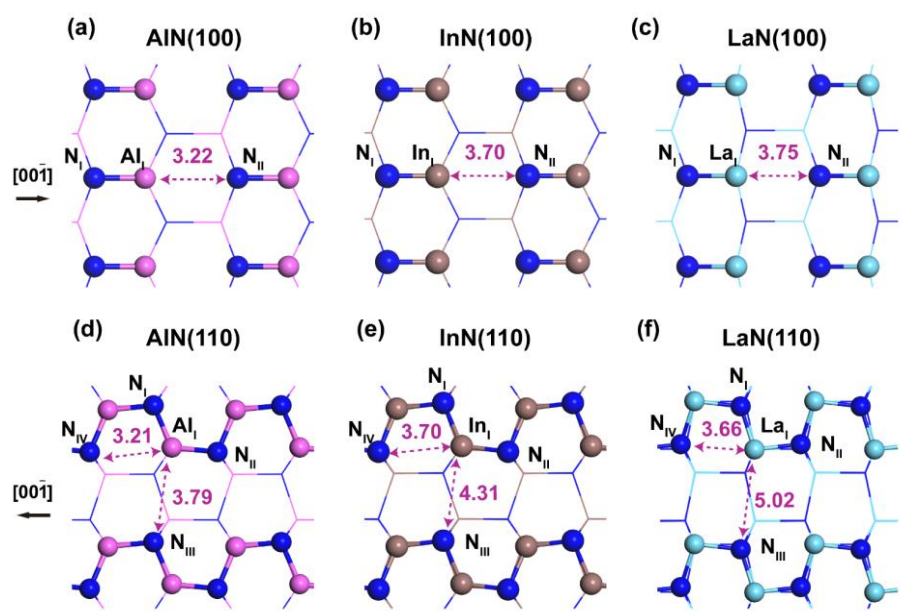

**Figure S10.** Optimized structures of (a) AlN(100), (b) InN(100), (c) LaN(100), (d) AlN(110), (e) InN(110) and (f) LaN(110). The natural SFLPs are labeled by dashed arrows. The distances between Lewis acid and Lewis base are in Å.

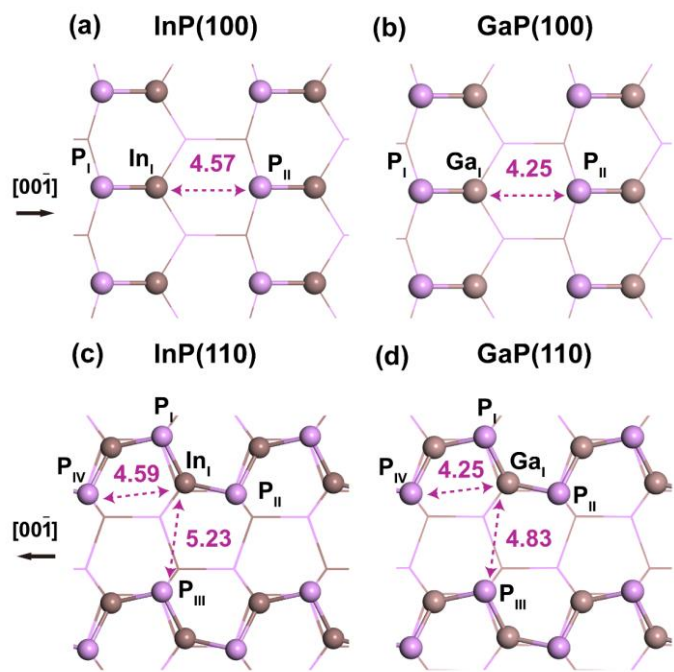

**Figure S11.** Optimized structures of (a) InP(100), (b) GaP(100), (c) InP(110), and (d) GaP(110). The natural SFLPs are labeled by dashed arrows. The distances between Lewis acid and Lewis base are in Å.

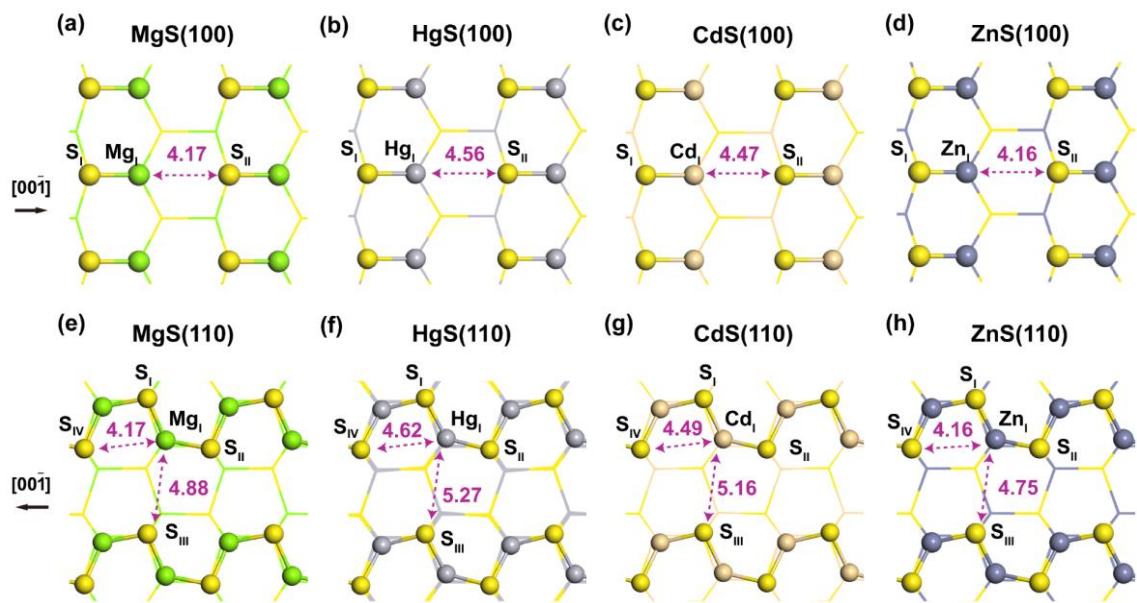

**Figure S12.** Optimized structures of (a) MgS(100), (b) HgS(100), (c) CdS(100), (d) ZnS(100), (e) MgS(110), (f) HgS(110), (g) CdS(110) and (h) ZnS(110). The natural SFLPs are labeled by dashed arrows. The distances between Lewis acid and Lewis base are in Å.

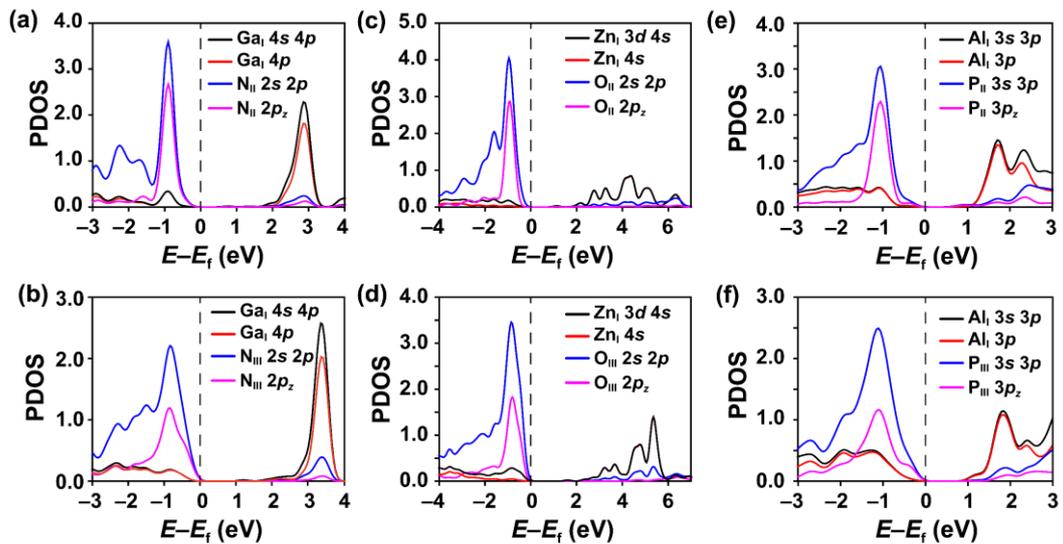

**Figure S13.** Frontier orbitals analysis of Lewis pairs on wurtzite crystal surfaces. PDOS of the selected atoms in the top atomic layer of (a) GaN(100), (b) GaN(110), (c) ZnO(100), (d) ZnO(110), (e) AlP(100), and (f) AlP(110).

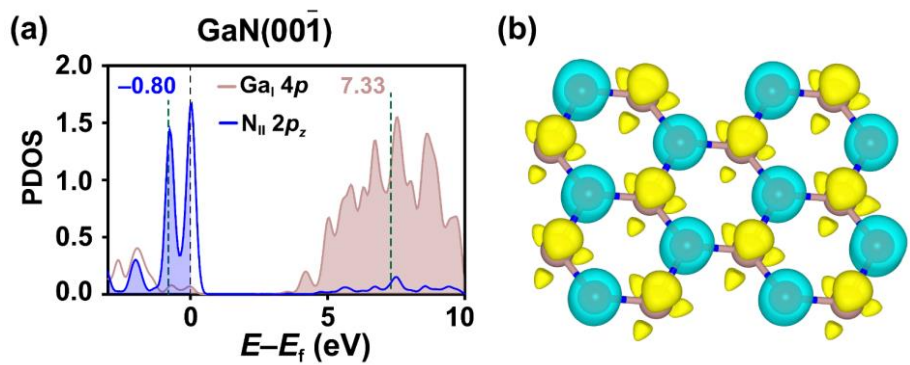

**Figure S14.** (a) Partial density of states (PDOS) of selected Ga and N atom on GaN (00$\bar{1}$). (b) Frontier orbitals maps of clean GaN(00$\bar{1}$). The energy-weighted band centers of N 2$p_z$ and Ga 4$p$ are calculated using the equation: $\varepsilon = (\sum \mathrm{PDOS}(E_i) \cdot E_i)/\sum \mathrm{PDOS}(E_i)$, where PDOS($E_i$) is the PDOS of a given orbital in an energy interval ($E_i$, $E_i + \Delta E$) and $\Delta E$ is 0.05 eV; $E_i$ of N 2$p_z$ orbitals, Ga 4$p$ orbitals are in a range from −3.0 to 0.0 eV, and 0.0 to 10.0 eV, respectively. The electron-density isosurfaces are plotted at 0.03 e/Bohr$^3$. The brown and blue balls represent Ga and N atoms, respectively. The yellow and cyan regions represent LUMO and HOMO, respectively.

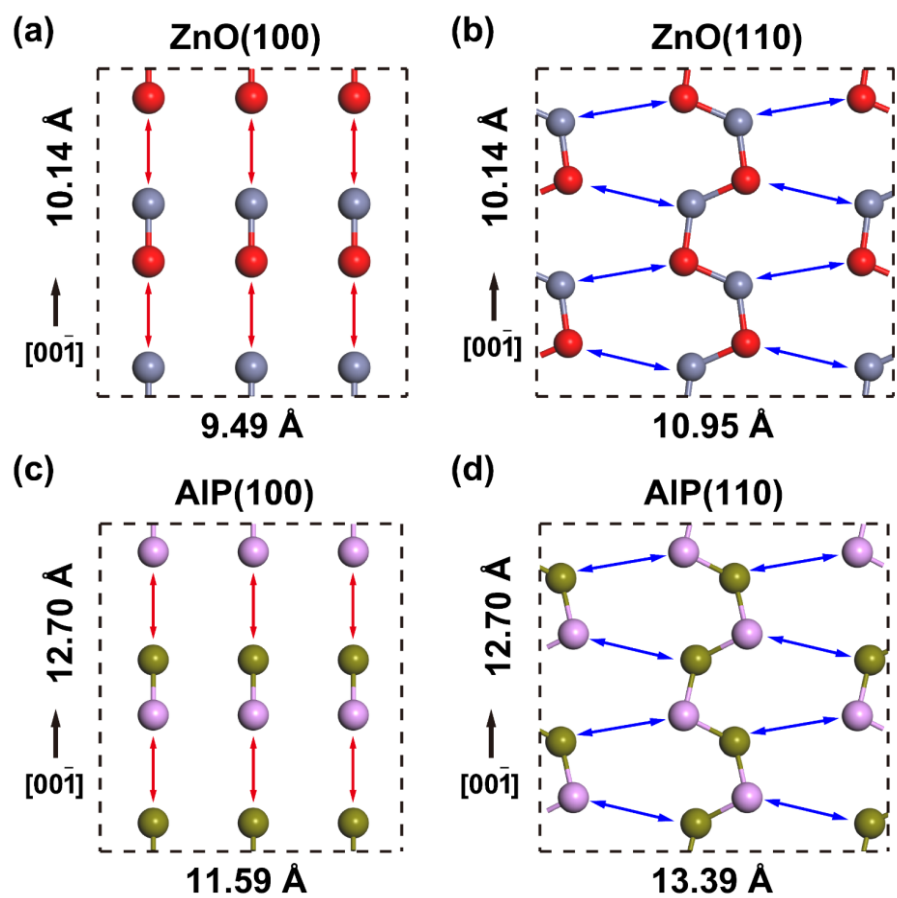

**Figure S15.** The distribution of SFLPs on (a) ZnO(100), (b) ZnO(110), (c) AlP(100), and (d) AlP(110) supercell.

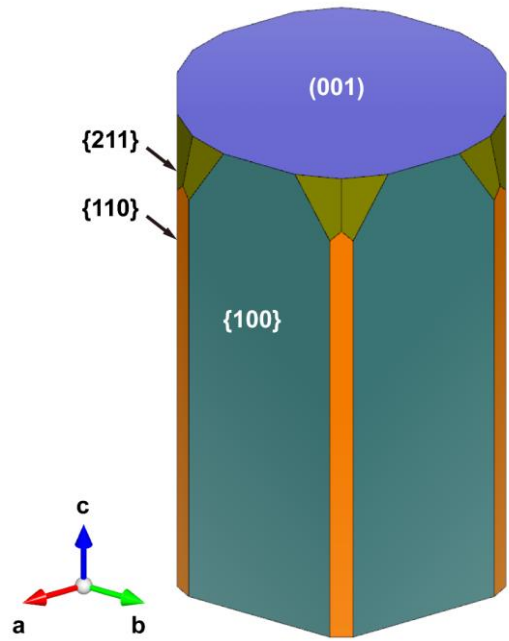

**Figure S16.** A 3D Wulff construction of GaN crystal.

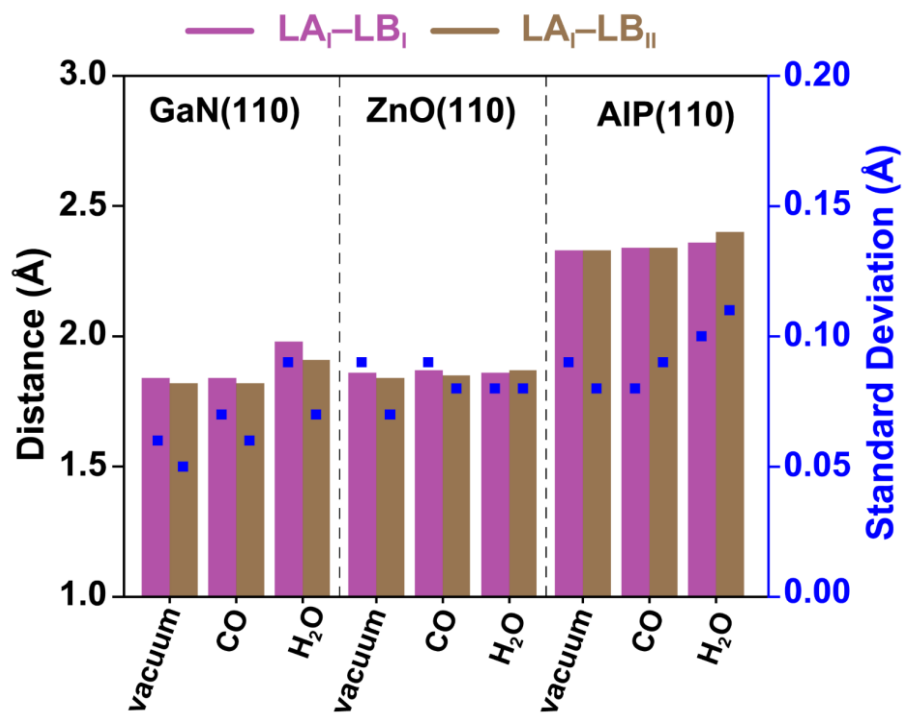

**Figure S17.** The mean distances and standard deviation of selected Lewis pairs on the (110) surfaces of wurtzite crystals during the AIMD simulations at 800 K with different atmospheres.

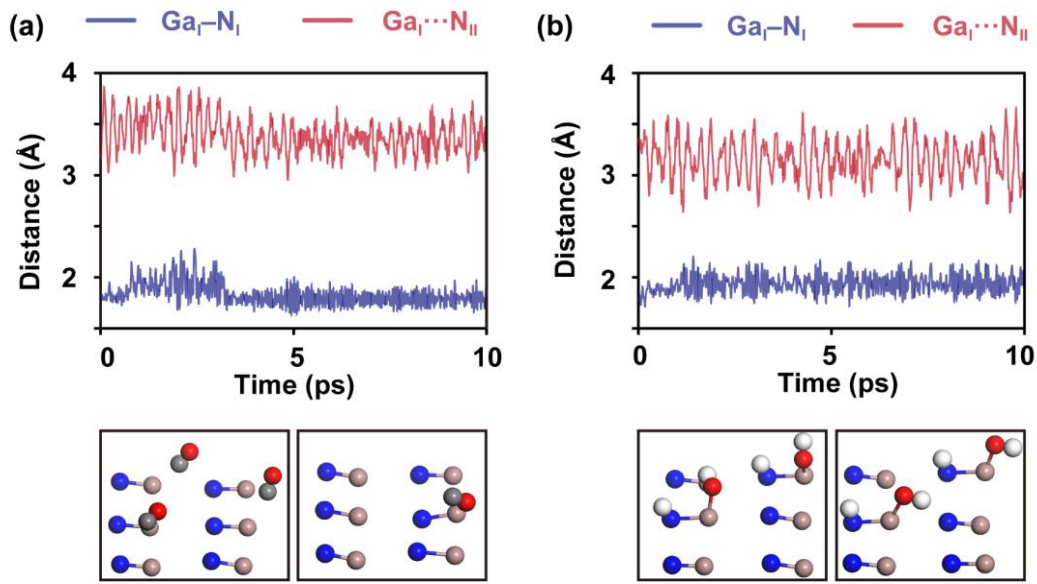

**Figure S18.** Stability analysis of natural SFLPs on GaN(100). Variation of the distances of selected Lewis pairs upon AIMD simulations with (a) CO and (b) $H_2O$ atmosphere on GaN(100) surface. The structures on the bottom side in (a-b) correspond to the states at 3.33 ps (left) and 6.67 ps (right). Some of molecules in the vacuum layer are not shown in the pictures. The brown, blue, gray, white, and red balls represent Ga, N, C, H, and O atoms, respectively.

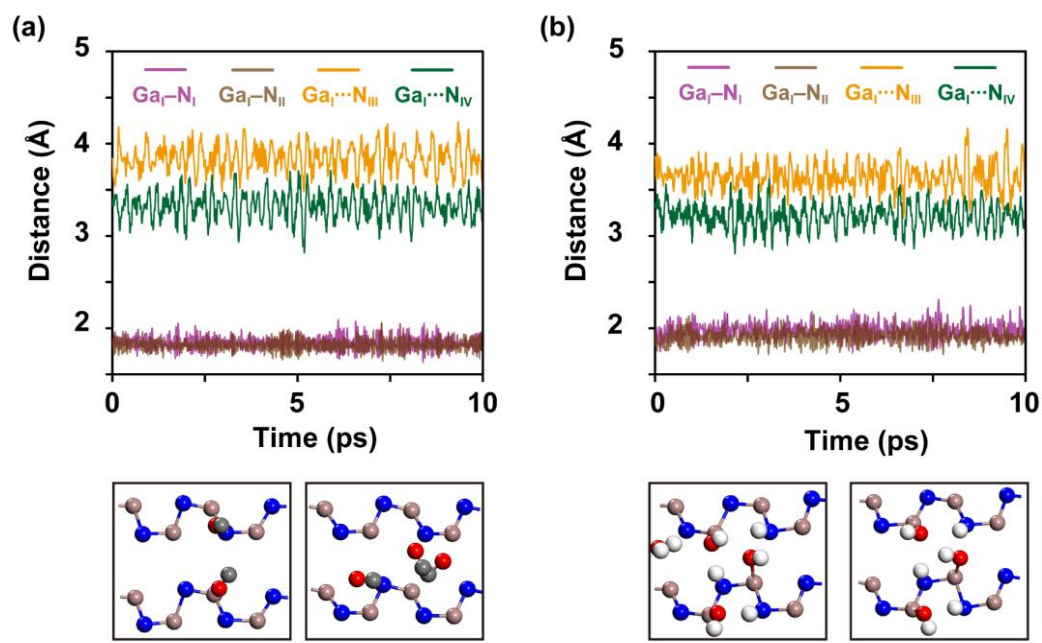

**Figure S19.** Stability analysis of natural SFLPs on GaN(110). Variation of the distances of selected Lewis pairs upon AIMD simulations with (a) CO and (b) $H_2O$ atmosphere on GaN(110) surface. The structures on the bottom side in (a-b) correspond to the states at 3.33 ps (left) and 6.67 ps (right). Some of molecules in the vacuum layer are not shown in the pictures. The brown, blue, gray, white, and red balls represent Ga, N, C, H, and O atoms, respectively.

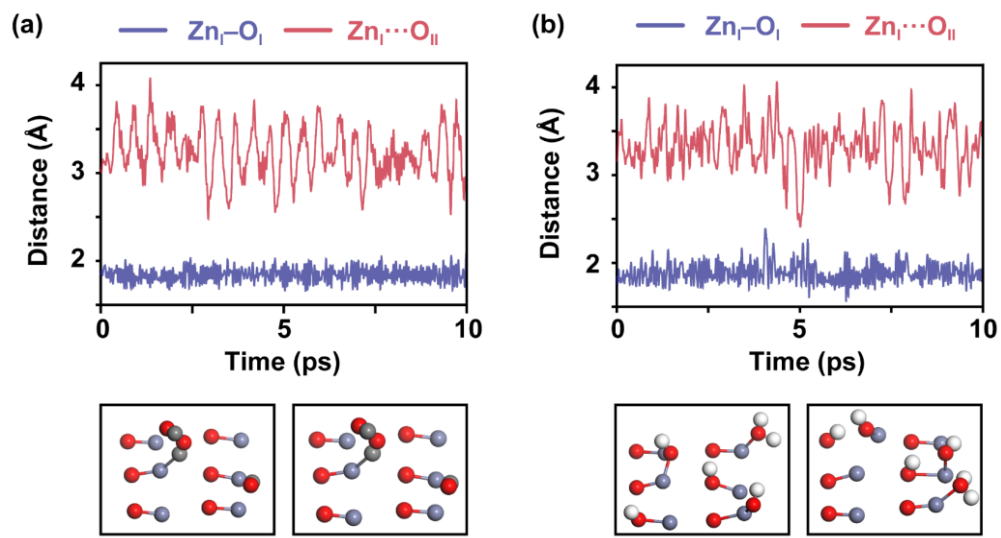

**Figure S20.** Stability analysis of natural SFLPs on ZnO(100). Variation of the distances of selected Lewis pairs upon AIMD simulations with (a) CO and (b) $H_2O$ atmosphere on ZnO(100) surface. The structures on the bottom side in (a-b) correspond to the states at 3.33 ps (left) and 6.67 ps (right). Some of molecules in the vacuum layer are not shown in the pictures. The shadow blue, red, gray, and white balls represent Zn, O, C, and H atoms, respectively.

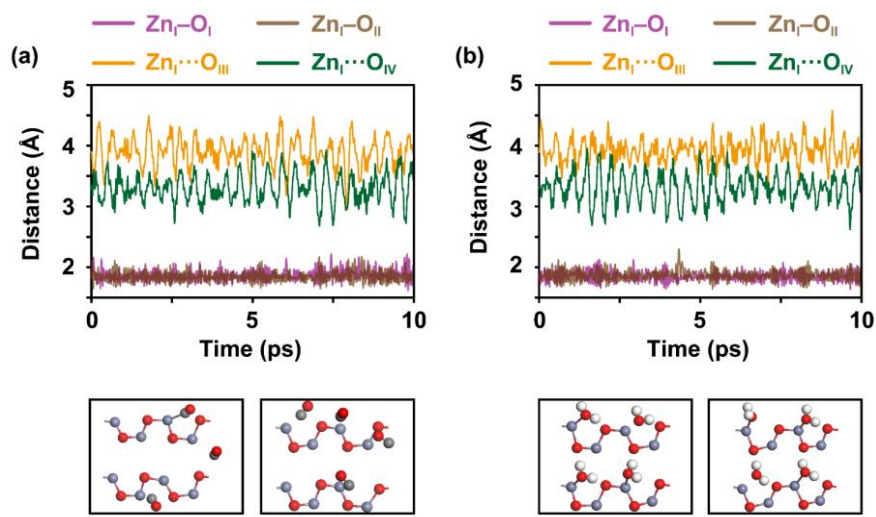

**Figure S21.** Stability analysis of natural SFLPs on ZnO(110). Variation of the distances of selected Lewis pairs upon AIMD simulations with (a) CO and (b) $H_2O$ atmosphere on ZnO(110) surface. The structures on the bottom side in (a-b) correspond to the states at 3.33 ps (left) and 6.67 ps (right). Some of molecules in the vacuum layer are not shown in the pictures. The shadow blue, red, gray, and white balls represent Zn, O, C, and H atoms, respectively.

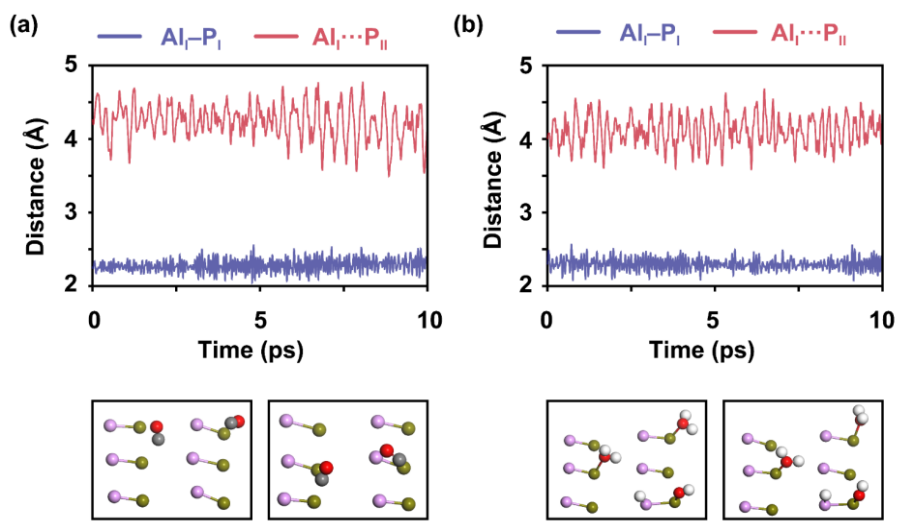

**Figure S22.** Stability analysis of natural SFLPs on AlP(100). Variation of the distances of selected Lewis pairs upon AIMD simulations with (a) CO and (b) H$_2$O atmosphere on AlP(100) surface. The structures on the bottom side in (a-b) correspond to the states at 3.33 ps (left) and 6.67 ps (right). Some of molecules in the vacuum layer are not shown in the pictures. The olive, pink, gray, red, and white balls represent Al, P, C, O, and H atoms, respectively.

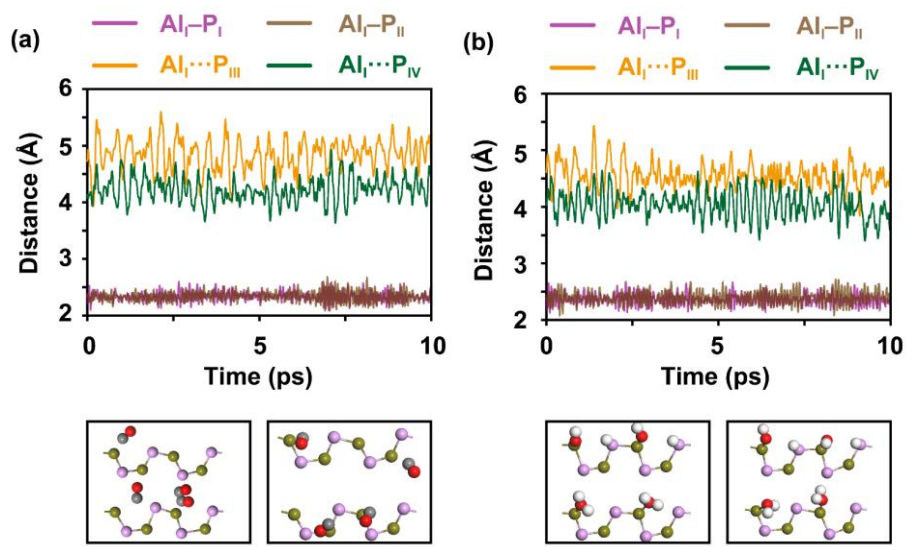

**Figure S23.** Stability analysis of natural SFLPs on AlP(110). Variation of the distances of selected Lewis pairs upon AIMD simulations with (a) CO and (b) $H_2O$ atmosphere on AlP(110) surface. The structures on the bottom side in (a-b) correspond to the states at 3.33 ps (left) and 6.67 ps (right). The olive, pink, gray, white, and red balls represent Al, P, C, H, and O atoms, respectively.

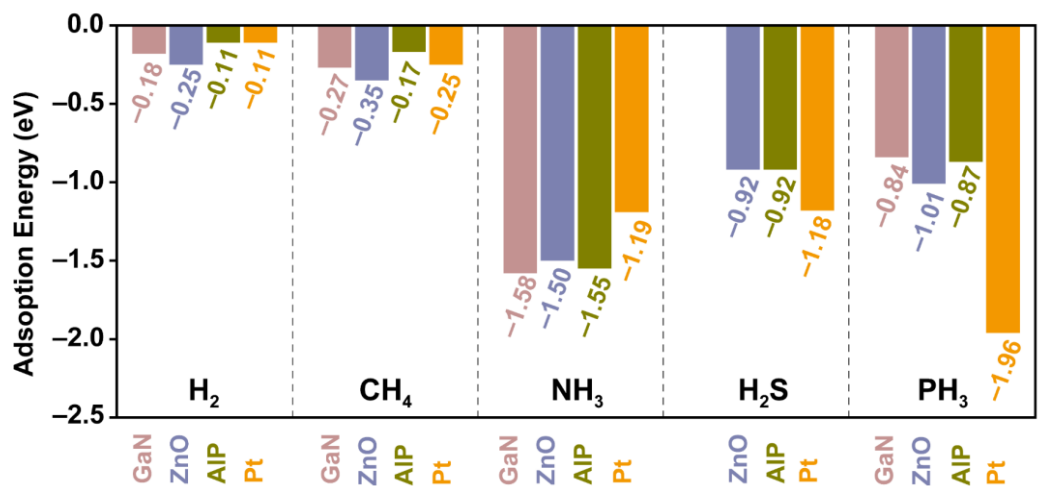

**Figure S24.** The adsorption energy of small molecules on SFLPs of wurtzite crystal surfaces and Pt(111).

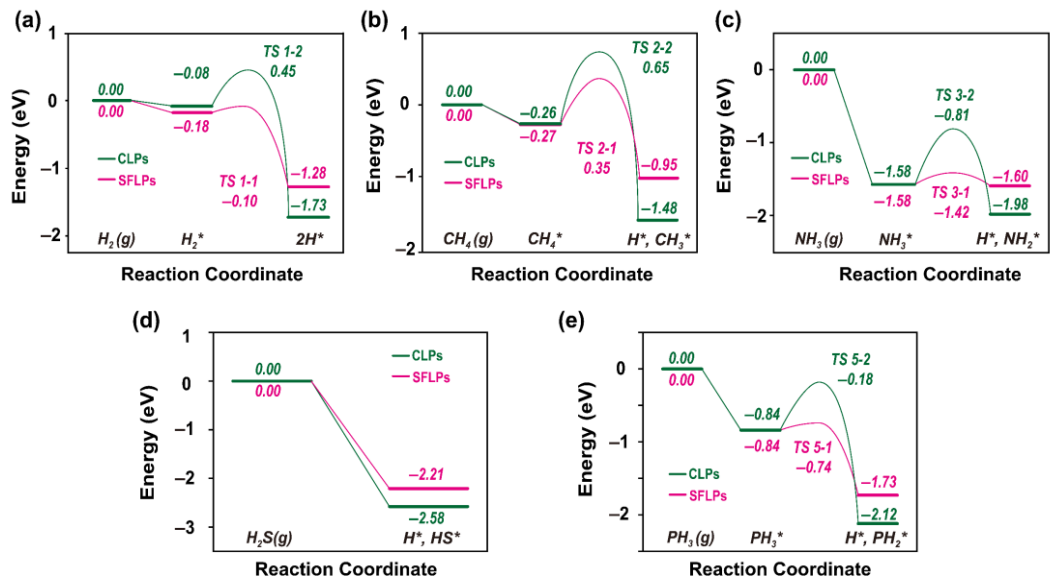

**Figure S25.** Potential energy profiles of (a) hydrogen, (b) methane, (c) ammonia, (d) hydrogen sulfide, and (e) phosphine dissociation on CLPs and SFLPs of GaN(100) surface.

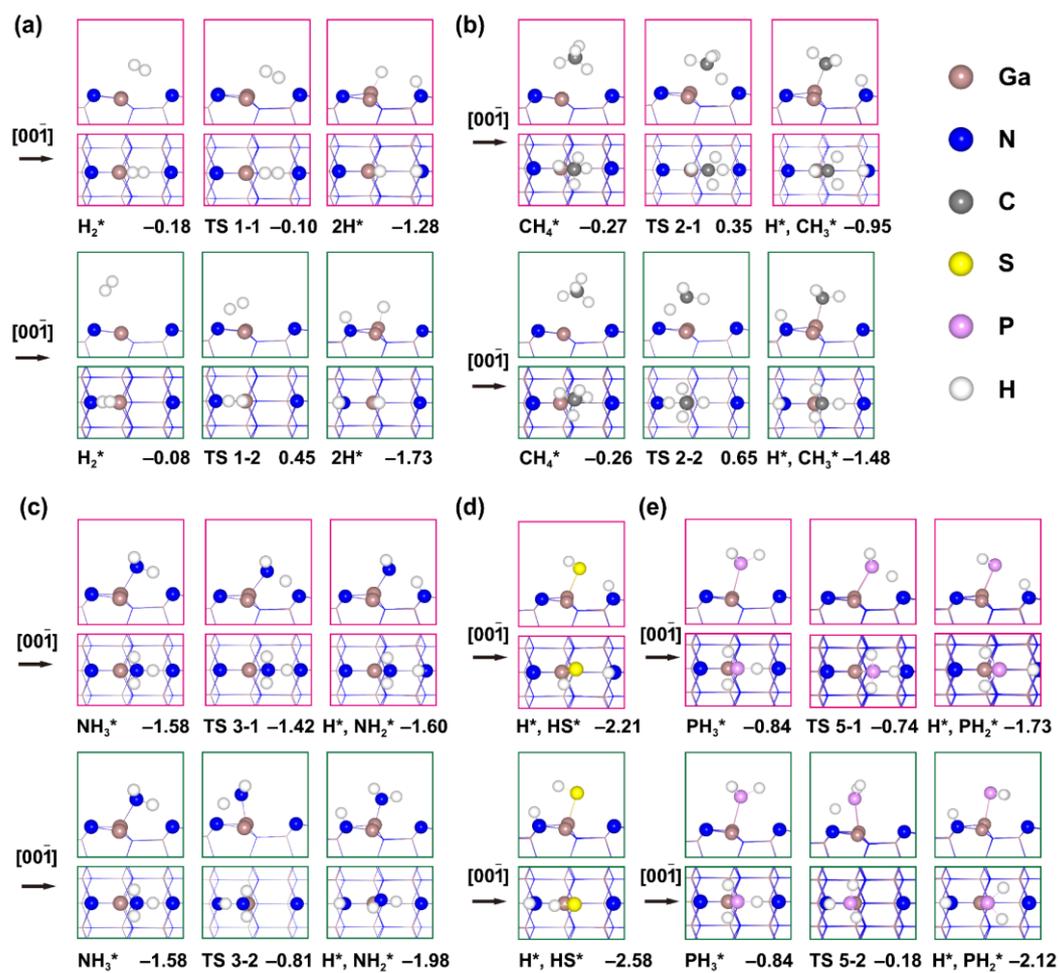

**Figure S26.** Side and top view structures of the corresponding initial state, transition state, and final state of (a) hydrogen, (b) methane, (c) ammonia, (d) hydrogen sulfide, and (e) phosphine dissociation on SFLPs and CLPs of GaN(100). The energetics labeled below the frame are in eV.

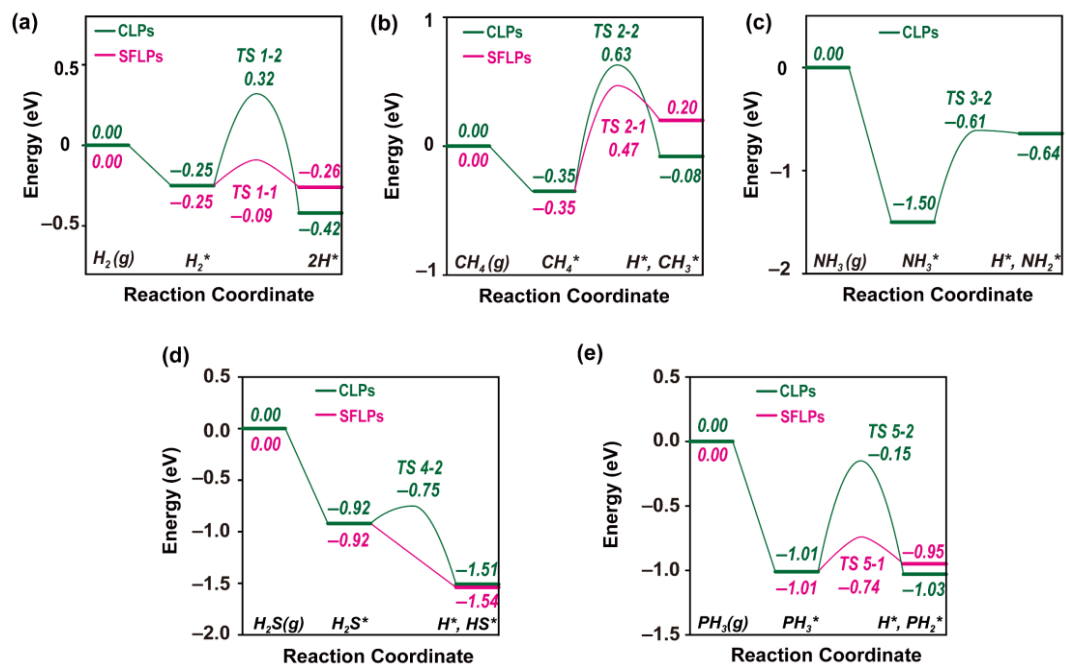

**Figure S27.** Potential energy profiles of (a) hydrogen, (b) methane, (c) ammonia, (d) hydrogen sulfide, and (e) phosphine dissociation on Lewis pairs of ZnO(100) surface.

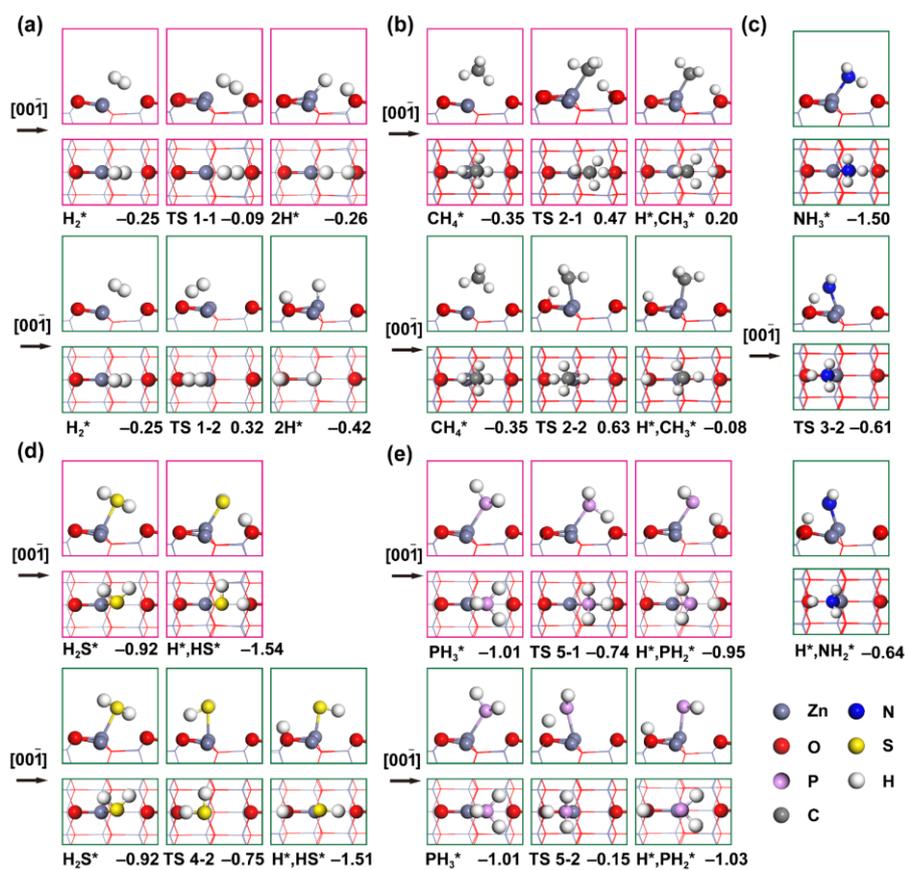

**Figure S28.** Side and top view structures of the corresponding initial state, transition state, and final state of (a) hydrogen, (b) methane, (c) ammonia, (d) hydrogen sulfide, and (e) phosphine dissociation on Lewis pairs of ZnO(100). The energetics labeled below the frame are in eV.

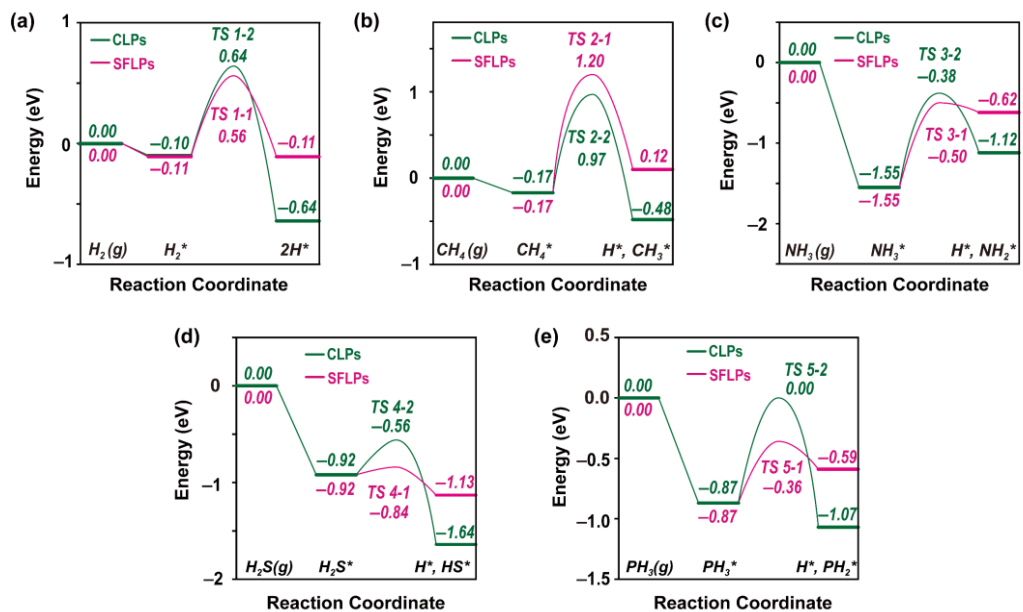

**Figure S29.** Potential energy profiles of (a) hydrogen, (b) methane, (c) ammonia, (d) hydrogen sulfide, and (e) phosphine dissociation on CLPs and SFLPs of AlP(100) surface.

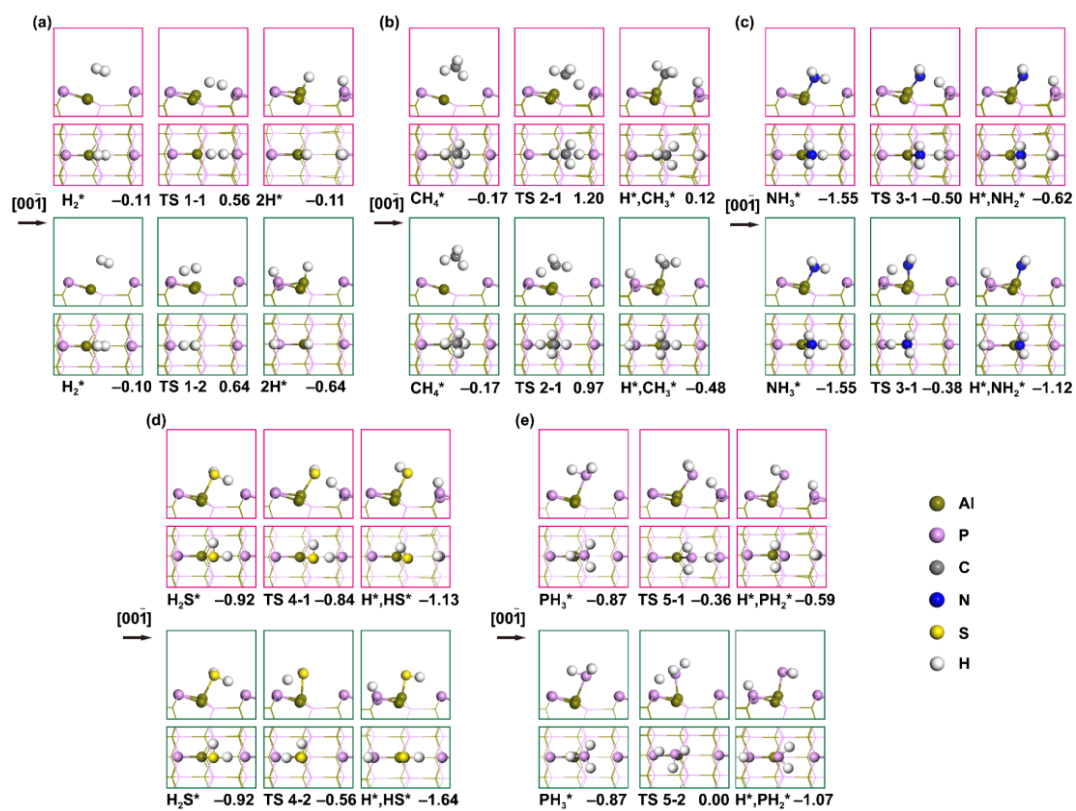

**Figure S30.** Side and top view structures of the corresponding initial state, transition state, and final state of (a) hydrogen, (b) methane, (c) ammonia, (d) hydrogen sulfide, and (e) phosphine dissociation on SFLPs and CLPs of AlP(100). The energetics labeled below the frame are in eV.

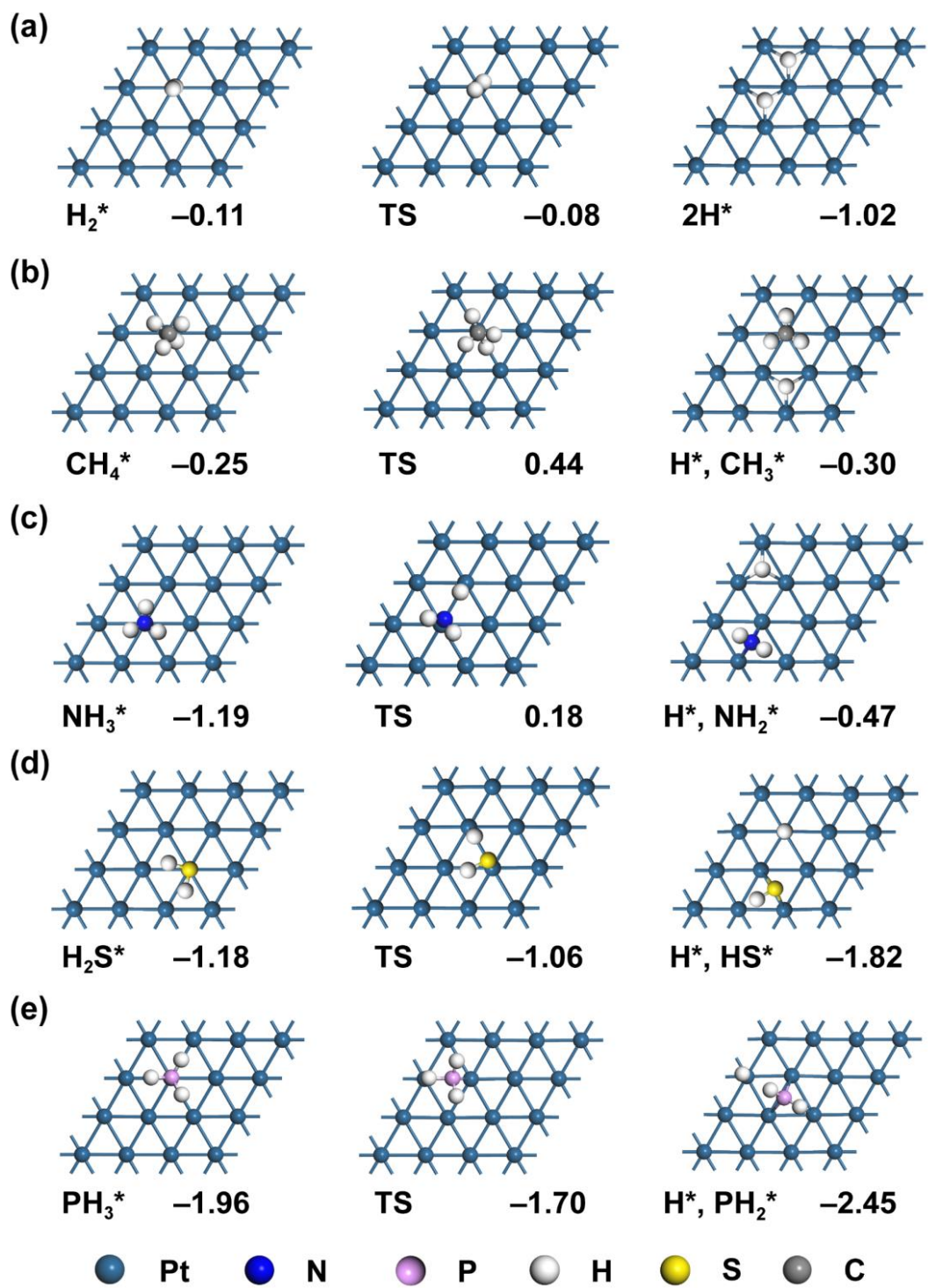

**Figure S31.** Optimized intermediates and transition states of small molecules dissociation on Pt(111). Top view structures of the corresponding initial state, transition state, and final state of (a) hydrogen, (b) methane, (c) ammonia, (d) hydrogen sulfide, and (e) phosphine dissociation. The energetics labeled below the pictures are in eV.

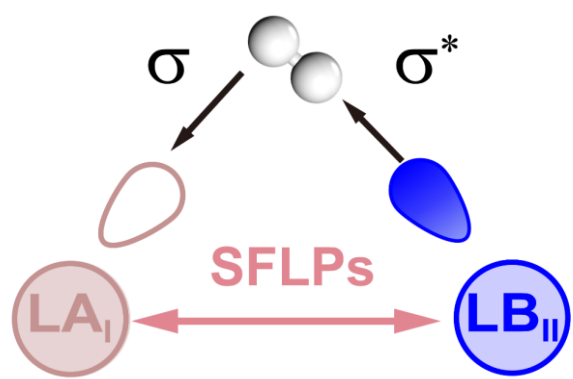

**Figure S32.** Schematic images of interactions between hydrogen and SFLPs.

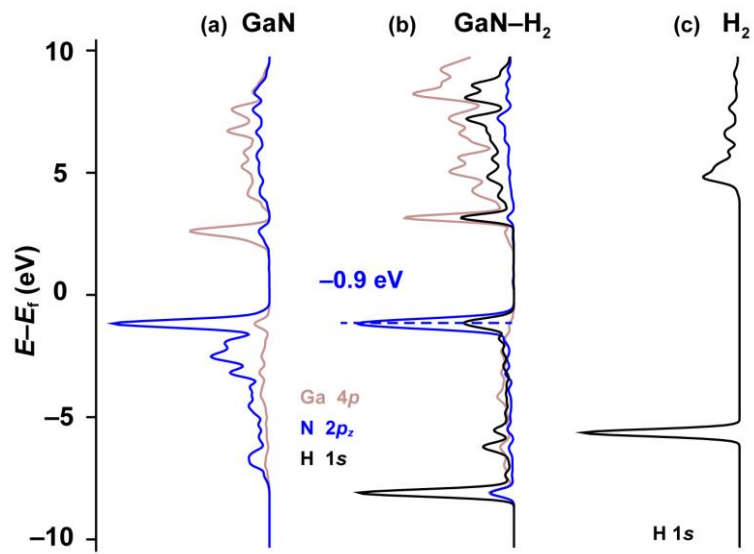

**Figure S33.** PDOS analysis of TS of hydrogen dissociation on CLPs of GaN(100). (a) The PDOS of surface Ga and N atoms on the clean surface. (b) PDOS of selected atoms of TS in hydrogen dissociation on CLPs. (c) 1s orbital for H atoms of the gaseous $H_2$. The Fermi levels of the GaN(100) surface are set to zero in this figure.

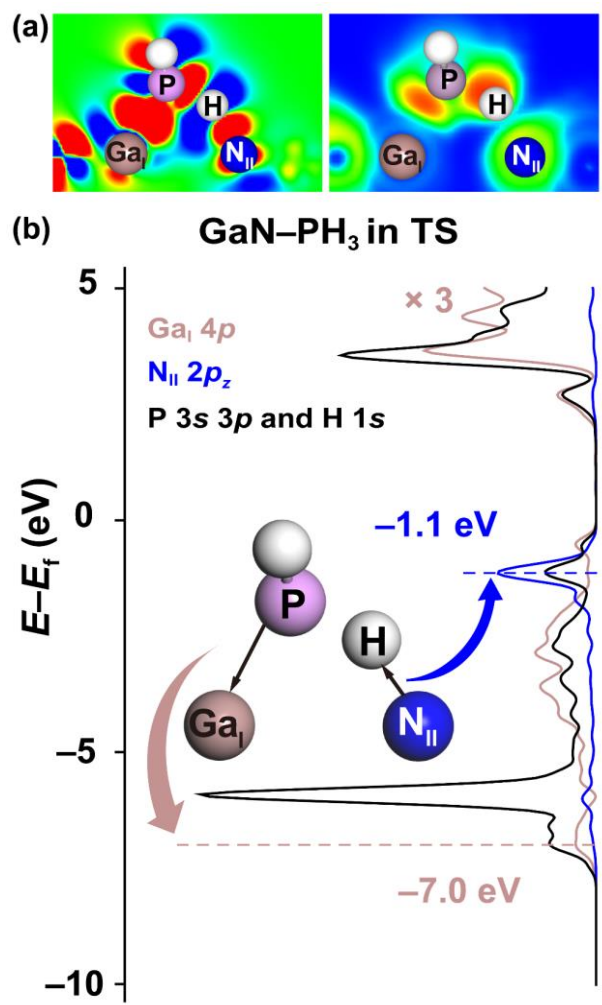

**Figure S34.** Electronic analysis of PH$_3$ activation on SFLP of GaN(100). (a) Charge density difference ($\Delta\rho = \rho$(TS) – $\rho$(surface[#]) – $\rho$(phosphine[#])) maps and their corresponding electron localization function map with a range from –0.002 to 0.002 e/Bohr$^3$, and 0.0 to 1.0, respectively. The atomic positions of surface[#] and phosphine[#] are identical to those in the TS. (b) PDOS analysis of the selected atoms of TS in PH$_3$ dissociation.

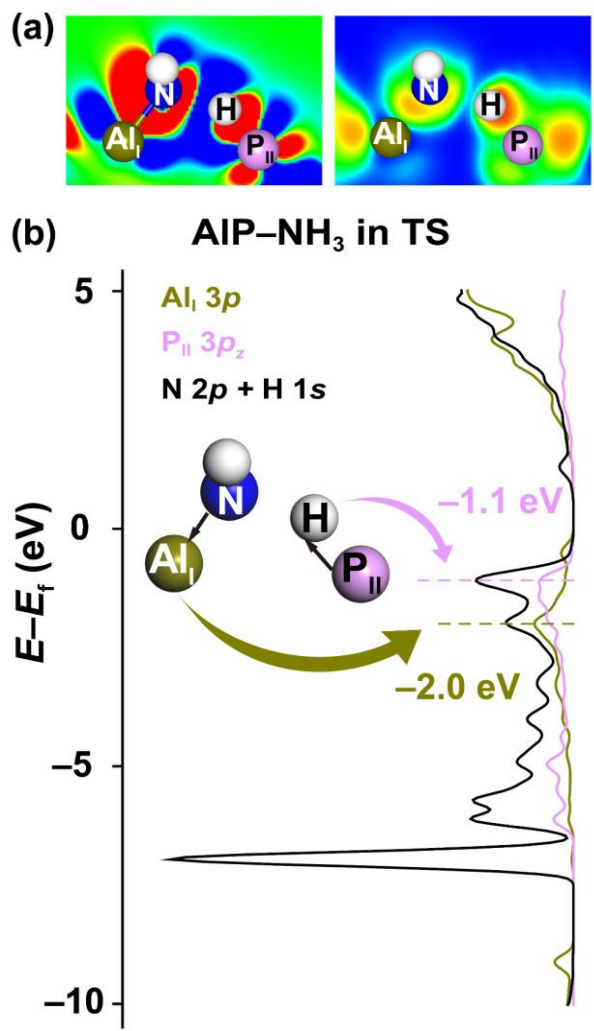

**Figure S35.** Electronic analysis of ammonia activation on SFLP of AlP(100). (a) Charge density difference ($\Delta\rho = \rho(TS) - \rho(\text{surface}^\#) - \rho(\text{ammonia}^\#)$) maps and their corresponding electron localization function map with a range from –0.002 to 0.002 e/Bohr$^3$, and 0.0 to 1.0, respectively. The atomic positions of surface$^\#$ and ammonia$^\#$ are identical to those in the TS. (b) PDOS analysis of the selected atoms of TS in NH$_3$ dissociation.

**Table S1.** The lattice parameters of various crystal structures for radial distribution function calculations.

| Structure type | Structure | lattice parameters | | |
|---|---|---|---|---|
| | | a (Å) | b (Å) | c (Å) |
| rocksalt | MgO | 4.26 | 4.26 | 4.26 |
| anti-fluorite | $Na_2O$ | 5.60 | 5.60 | 5.60 |
| rutile | $TiO_2$ | 4.65 | 4.65 | 2.97 |
| zincblence | ZnS | 5.45 | 5.45 | 5.45 |
| fluorite | $CeO_2$ | 5.47 | 5.47 | 5.47 |
| corundum | $Al_2O_3$ | 4.81 | 4.81 | 13.12 |
| anatase | $TiO_2$ | 3.80 | 3.80 | 9.75 |
| cuprite | $Cu_2O$ | 4.29 | 4.29 | 4.29 |
| CuO | CuO | 4.69 | 3.42 | 5.13 |
| quartz | $GeO_2$ | 5.09 | 5.09 | 5.76 |
| bixbyite | $In_2O_3$ | 8.92 | 8.92 | 8.92 |
| A-type $M_2O_3$ | $La_2O_3$ | 3.94 | 3.94 | 6.18 |
| B-type $M_2O_3$ | $Ce_2O_3$ | 14.44 | 3.70 | 8.98 |

**Table S2.** The calculated lattice parameters of wurtzite crystal structures.

[a] referred from [17a]

| structures | Calculated Lattice Parameter | | |
|:---:|:---:|:---:|:---:|
| | a (Å) | b (Å) | c (Å) |
| GaN [a] | 3.12 | 3.12 | 5.09 |
| ZnO [b] | 3.16 | 3.16 | 5.07 |
| AlP | 3.86 | 3.86 | 6.35 |
| CdO | 3.65 | 3.65 | 5.82 |
| BaO | 4.32 | 4.32 | 6.39 |
| BeO | 2.68 | 2.68 | 4.35 |
| AlN | 3.10 | 3.10 | 4.97 |
| InN | 3.55 | 3.55 | 5.73 |
| LaN | 4.11 | 4.11 | 5.92 |
| InP | 4.18 | 4.18 | 6.86 |
| GaP | 3.85 | 3.85 | 6.36 |
| MgS | 3.99 | 3.99 | 6.39 |
| HgS | 4.18 | 4.18 | 6.84 |
| CdS | 4.15 | 4.15 | 6.75 |
| ZnS | 3.80 | 3.80 | 6.23 |

[b] referred from [17b]

**Table S3.** Cartesian coordinates of Lewis pairs on GaN(100) and the centroid of its corresponding frontier orbitals.

| Atom | Coordinate | Centriods[a] | Coordinate |
|---|---|---|---|
| $Ga_I$ | (4.69, 7.85, 6.13) | $Ga_I$' | (4.69, 7.45, 7.08) |
| $Ga_{II}$ | (4.69, 2.76, 6.13) | $Ga_{II}$' | (4.69, 2.36, 7.08) |
| $N_I$ | (4.69, 9.62, 6.35) | $N_I$' | (4.69, 9.81, 6.84) |
| $N_{II}$ | (4.69, 4.53, 6.35) | $N_{II}$' | (4.69, 4.72, 6.84) |

[a] The charge density is obtained from the PARCHG file, the calculated formula of the centroid is $x = (\sum \rho(X_i) \cdot X_i)/\sum \rho(X_i)$. The lattce parameter is a = 9.37 Å, b = 10.18 Å, and c = 21.31 Å, respectively.

**Table S4.** Selected frontier orbital parameters on GaN(100) surface.

| Parameter | Angle/Distance |
|---|---|
| ∠**A₁, Ga$_I$–N$_I$** | 105.7° |
| ∠**B₁, N$_I$–Ga$_I$** | 118.3° |
| ∠**B₂, N$_{II}$–Ga$_{II}$** | 118.3° |
| d(Ga$_I$', N$_{II}$') | 2.74 Å |
| d(Ga$_I$', N$_I$') | 2.37 Å |

**Table S5**. The mean distance and the standard deviation of the Ga–N and Ga···N Lewis pairs on GaN(100) in AIMD simulations at 800 K.

| Lewis Pair | Vacuum | | CO | | $H_2O$ | |
|---|---|---|---|---|---|---|
| | Standard Deviation (Å) | Mean (Å) | Standard Deviation (Å) | Mean (Å) | Standard Deviation (Å) | Mean (Å) |
| $Ga_I$–$N_I$ | 0.05 | 1.79 | 0.10 | 1.83 | 0.08 | 1.94 |
| $Ga_I$···$N_{II}$ | 0.13 | 3.33 | 0.17 | 3.39 | 0.19 | 3.15 |

**Table S6**. The mean distance and the standard deviation of the Ga–N and Ga···N Lewis pairs on GaN(110) in AIMD simulations at 800 K.

| Lewis Pair | Vacuum | | CO | | $H_2O$ | |
|---|---|---|---|---|---|---|
| | Standard Deviation (Å) | Mean (Å) | Standard Deviation (Å) | Mean (Å) | Standard Deviation (Å) | Mean (Å) |
| $Ga_I$–$N_I$ | 0.06 | 1.84 | 0.07 | 1.84 | 0.09 | 1.98 |
| $Ga_I$–$N_{II}$ | 0.05 | 1.82 | 0.06 | 1.82 | 0.07 | 1.91 |
| $Ga_I$···$N_{III}$ | 0.17 | 3.85 | 0.14 | 3.84 | 0.14 | 3.65 |
| $Ga_I$···$N_{IV}$ | 0.18 | 3.33 | 0.14 | 3.32 | 0.13 | 3.21 |

**Table S7.** The mean distance and the standard deviation of the Zn–O and Zn···O Lewis pairs on ZnO(100) in AIMD simulations at 800 K.

| Lewis Pair | Vacuum | | CO | | $H_2O$ | |
|---|---|---|---|---|---|---|
| | Standard Deviation (Å) | Mean (Å) | Standard Deviation (Å) | Mean (Å) | Standard Deviation (Å) | Mean (Å) |
| $Zn_I$–$O_I$ | 0.08 | 1.83 | 0.07 | 1.85 | 0.11 | 1.89 |
| $Zn_I$···$O_{II}$ | 0.24 | 3.32 | 0.28 | 3.23 | 0.26 | 3.32 |

**Table S8.** The mean distance and the standard deviation of the Zn–O and Zn⋯O Lewis pairs on ZnO(110) in AIMD simulations at 800 K.

| Lewis Pair | Vacuum | | CO | | $H_2O$ | |
|---|---|---|---|---|---|---|
| | Standard Deviation (Å) | Mean (Å) | Standard Deviation (Å) | Mean (Å) | Standard Deviation (Å) | Mean (Å) |
| $Zn_{II}$–$O_I$ | 0.09 | 1.87 | 0.09 | 1.87 | 0.08 | 1.86 |
| $Zn_{II}$–$O_{II}$ | 0.07 | 1.84 | 0.08 | 1.85 | 0.08 | 1.87 |
| $Zn_I$⋯$O_{III}$ | 0.25 | 3.88 | 0.24 | 3.89 | 0.18 | 3.92 |
| $Zn_I$⋯$O_{IV}$ | 0.18 | 3.30 | 0.23 | 3.30 | 0.25 | 3.30 |

**Table S9.** The mean distance and the standard deviation of the Al–P and Al⋯P Lewis pairs on AlP(100) in AIMD simulations at 800 K.

| Lewis Pair | Vacuum | | CO | | H$_2$O | |
|---|---|---|---|---|---|---|
| | Standard Deviation (Å) | Mean (Å) | Standard Deviation (Å) | Mean (Å) | Standard Deviation (Å) | Mean (Å) |
| Al$_I$–P$_I$ | 0.07 | 2.27 | 0.08 | 2.28 | 0.08 | 2.30 |
| Al$_I$⋯P$_{II}$ | 0.28 | 4.28 | 0.24 | 4.23 | 0.20 | 4.11 |

**Table S10**. The mean distance and the standard deviation of the Al–P and Al⋯P Lewis pairs on AlP(110) in AIMD simulations at 800 K.

| Lewis Pair | Vacuum | | CO | | $H_2O$ | |
|---|---|---|---|---|---|---|
| | Standard Deviation (Å) | Mean (Å) | Standard Deviation (Å) | Mean (Å) | Standard Deviation (Å) | Mean (Å) |
| $Al_I$–$P_I$ | 0.09 | 2.33 | 0.08 | 2.34 | 0.10 | 2.36 |
| $Al_I$–$P_{II}$ | 0.08 | 2.33 | 0.09 | 2.34 | 0.11 | 2.40 |
| $Al_I$⋯$P_{III}$ | 0.29 | 4.90 | 0.28 | 4.84 | 0.22 | 4.55 |
| $Al_I$⋯$P_{IV}$ | 0.21 | 4.30 | 0.22 | 4.22 | 0.24 | 4.04 |

**Table S11.** Selected bond lengths (Å) and bond angles (°) of TS in dissociation of $H_2$, $CH_4$, $NH_3$, and $PH_3$ on SFLPs of GaN(100).

| Species | Sites | $d$(X–H)[a] | $d$(Ga$_I$–X) | $d$(N$_{II}$–H) | ∠HXGa$_I$ | ∠XHN$_{II}$ |
|---|---|---|---|---|---|---|
| $H_2$ | CLPs | 0.956[b] | 1.868[b] | 1.549[b] | 70.5 | 146.8 |
| | SFLPs | 0.877 | 1.931 | 1.672 | 100.8 | 165.9 |
| $CH_4$ | CLPs | 1.379 | 2.183 | 1.454 | 54.3 | 157.1 |
| | SFLPs | 1.360 | 2.207 | 1.416 | 74.0 | 174.6 |
| $NH_3$ | CLPs | 1.280 | 1.963 | 1.398 | 71.5 | 136.8 |
| | SFLPs | 1.349 | 1.950 | 1.288 | 98.5 | 159.9 |
| $PH_3$ | CLPs | 1.560 | 2.416 | 1.642 | 69.3 | 124.7 |
| | SFLPs | 1.519 | 2.426 | 1.666 | 93.9 | 143.9 |

[a] X = H, C, N, and P atoms

[b] The data are referred from [17b] for comparison

**Table S12.** Selected bond lengths (Å) and bond angles (°) of TS in dissociation of $H_2$, $CH_4$, and $PH_3$ on SFLPs of ZnO(100).

| Species | Sites | $d$(X–H)[a] | $d$(Zn–X) | $d$(O$_{II}$–H) | ∠HXZn$_I$ | ∠XHO$_{II}$ |
|---|---|---|---|---|---|---|
| $H_2$ | CLPs | 0.965 | 1.784 | 1.367 | 72.9 | 153.9 |
|  | SFLPs | 0.995 | 1.728 | 1.300 | 104.1 | 168.9 |
| $CH_4$ | CLPs | 1.438 | 2.117 | 1.254 | 55.8 | 165.2 |
|  | SFLPs | 1.487 | 2.108 | 1.188 | 73.7 | 174.9 |
| $PH_3$ | CLPs | 1.596 | 2.351 | 1.419 | 67.2 | 132.7 |
|  | SFLPs | 1.592 | 2.340 | 1.381 | 93.3 | 151.6 |

[a] X = H, C, and P atoms

**Table S13.** Selected bond lengths (Å) and bond angles (°) of TS in dissociation of $NH_3$, $H_2S$, and $PH_3$ on SFLPs of AlP(100).

| Species | Sites | $d(X–H)$[a] | $d(Al_I–X)$ | $d(P_{II}–H)$ | ∠$HXAl_I$ | ∠$XHP_{II}$ |
|---|---|---|---|---|---|---|
| $NH_3$ | CLPs | 1.553 | 1.891 | 1.625 | 75.3 | 136.9 |
|  | SFLPs | 1.841 | 1.882 | 1.512 | 110.2 | 146.7 |
| $H_2S$ | CLPs | 1.586 | 2.415 | 1.876 | 64.2 | 144.7 |
|  | SFLPs | 1.708 | 2.382 | 1.693 | 95.1 | 161.7 |
| $PH_3$ | CLPs | 1.643 | 2.411 | 1.840 | 78.4 | 122.9 |
|  | SFLPs | 1.888 | 2.405 | 1.612 | 103.5 | 147.6 |

[a] X = N, S, and P atoms

## Author Contributions

C.-R.C. and Z.-W.L. designed and supervised the project and wrote the paper. X.-Y.Y. performed all of the DFT calculations and co-wrote the paper. Z.-Q.H., T.B., and Y.-H.X. analyzed the data. All of the authors discussed the results of the paper.